\documentclass[12pt]{article}
\usepackage{a4p}
\usepackage{cite}
\usepackage{epsfig}
\usepackage{amsmath}
\usepackage{amssymb}
\usepackage{amsfonts}
\usepackage{latexsym}

\newcommand{\inmath}[1] {\ifmmode#1\else$#1$\fi}
\newcommand{\definmath}[2] {\def#1{\ifmmode#2\else$#2$\fi}}

\definmath{\Pq}      {\mathrm{q}}
\definmath{\Paq}  {\overline{\mathrm{q}}}

\newcommand {\ra}      {\rightarrow}
\newcommand {\LL}        {{\mathrm L^+ L^-}}

\newcommand {\lstar}  {\ell^{*}}
\newcommand {\estar}  {{\mathrm e}^{*}}
\newcommand {\mstar}  {\mu^{*}}
\newcommand {\tstar}  {\tau^{*}}

\newcommand {\nustarm} {\nu_{\mu}^{*}}
\newcommand {\nustart} {\nu_{\tau}^{*}}

\newcommand {\nul}     {\rm N_{\rm L}}

\newcommand{\fp} {f^{\prime}}

\begin{document}

%=======================================================================

\begin{titlepage}
\begin{center}{\large  EUROPEAN ORGANISATION FOR NUCLEAR RESEARCH}
\end{center}\bigskip
\begin{flushright} CERN-EP/99-169\\ 29 November 1999
\end{flushright}

\bigskip\bigskip\bigskip\bigskip\bigskip
\begin{center}{\huge\bf\boldmath
    Search for Unstable Heavy and Excited Leptons at LEP2
}\end{center}

\bigskip\bigskip
\begin{center}{\LARGE The OPAL Collaboration}\end{center}
\bigskip\bigskip

\bigskip\begin{center}{\large  Abstract}\end{center}
Searches for unstable neutral and charged
heavy leptons, N and ${\rm L^\pm}$, and
for excited states of neutral and charged leptons,
$\nu^*$, ${\rm e}^*$, $\mu^*$, and $\tau^*$,
have been performed in ${\rm e^+e^-}$ collisions using 
data collected by the OPAL detector at LEP.
The data analysed correspond to an integrated luminosity of about
58~pb$^{-1}$ at a centre-of-mass energy of 183~GeV, and 
about 10~pb$^{-1}$ each at 161~GeV and 172~GeV.
No evidence for new particles was found.
Lower limits on the masses of unstable heavy and excited leptons 
are derived.
From the analysis of charged-current, neutral-current, and photonic decays
of singly produced excited leptons,
upper limits are determined for
the ratio of the coupling to the compositeness scale,
$f/\Lambda$,  for masses up to
the kinematic limit.  For excited leptons, the limits are established 
independently of the relative values
of the coupling constants $f$ and $f^\prime$.
\bigskip\bigskip\bigskip\bigskip
\bigskip\bigskip
\begin{center}{\large
(submitted to Eur.~Phys.~J.~C)
}\end{center}

\end{titlepage}

%============================================================================
\begin{center}{
%begin authorlist PLEASE DO NOT DELETE THIS COMMENT
G.\thinspace Abbiendi$^{  2}$,
K.\thinspace Ackerstaff$^{  8}$,
P.F.\thinspace Akesson$^{  3}$,
G.\thinspace Alexander$^{ 23}$,
J.\thinspace Allison$^{ 16}$,
K.J.\thinspace Anderson$^{  9}$,
S.\thinspace Arcelli$^{ 17}$,
S.\thinspace Asai$^{ 24}$,
S.F.\thinspace Ashby$^{  1}$,
D.\thinspace Axen$^{ 29}$
% ...
%end authorlist PLEASE DO NOT DELETE THIS COMMENT
}\end{center}\bigskip
\bigskip
%begin institutes
$^{  1}$School of Physics and Astronomy, University of Birmingham,
Birmingham B15 2TT, UK
\newline
$^{  2}$Dipartimento di Fisica dell' Universit\`a di Bologna and INFN,
I-40126 Bologna, Italy
\newline
$^{  3}$Physikalisches Institut, Universit\"at Bonn,
D-53115 Bonn, Germany
\newline
% ...
$^{  8}$CERN, European Organisation for Particle Physics,
CH-1211 Geneva 23, Switzerland
\newline
$^{  9}$Enrico Fermi Institute and Department of Physics,
University of Chicago, Chicago IL 60637, USA
\newline
% ...
$^{ 16}$Department of Physics, Schuster Laboratory, The University,
Manchester M13 9PL, UK
\newline
$^{ 17}$Department of Physics, University of Maryland,
College Park, MD 20742, USA
\newline
% ...
$^{ 23}$Department of Physics and Astronomy, Tel Aviv University,
Tel Aviv 69978, Israel
\newline
$^{ 24}$International Centre for Elementary Particle Physics and
Department of Physics, University of Tokyo, Tokyo 113-0033, and
Kobe University, Kobe 657-8501, Japan
\newline
% ...
$^{ 29}$University of British Columbia, Department of Physics,
Vancouver BC V6T 1Z1, Canada
\newline
% ...

%end notes

\newpage

%============================================================================
\section{Introduction}
\label{s:intro}

In spite of its remarkable success in describing 
all electroweak data available today, the Standard Model (SM)~\cite{ref:sm}
leaves many questions unanswered. In particular, it explains neither
the origin of the number of fermion generations nor the fermion mass spectrum.
The precise measurements of the electroweak parameters
at the Z pole have shown that the number of species of
light neutrinos is three~\cite{ref:pdg};
however, this does not exclude a fourth generation,
or other massive fermions,
if these particles have masses greater than half the mass 
of the Z-boson ($M_{\rm Z}/2$).

New fermions could be of the following 
types (for reviews see References~\cite{ref:revue,ref:abdel,ref:bdk}):
sequential fermions,
mirror fermions (with chirality opposite to that in the SM),
vector fermions (with left- and right-handed doublets), and
singlet fermions.
These could be produced at high-energy ${\rm e^+e^-}$ colliders
such as LEP,
where two production mechanisms are possible: pair-production
and single-production in association with a light standard
fermion.

Lower limits on the masses of heavy leptons were obtained 
in ${\rm e^+e^-}$ collisions at centre-of-mass energies,
$\sqrt{s}$, around $ M_{\rm Z}$ \cite{ref:pdg,ref:hllep1}, and
recent searches at $\sqrt{s}=$~130-140~GeV \cite{ref:hlOPAL15,ref:hllep15},
$\sqrt{s} =$~161~GeV \cite{ref:hlOPAL161,ref:hlL3172}, 
$\sqrt{s} =$~172~GeV 
\cite{ref:hlL3172,ref:hlOPAL172}
and $\sqrt{s} =$~130-183~GeV \cite{ref:stableOPAL183,ref:DELPHI183}
have improved these limits.
Excited leptons have been sought at
$\sqrt{s} \sim M_{\rm Z}$
\cite{ref:ellep1},
$\sqrt{s}=$~130-140~GeV
\cite{ref:elopal15,ref:ellep15},
$\sqrt{s}=$~161~GeV \cite{ref:elopal161,ref:ellep161},
$\sqrt{s}=$~172~GeV \cite{ref:hlOPAL172},
$\sqrt{s}=$~183~GeV \cite{ref:DELPHI183},
$\sqrt{s}=$~189~GeV \cite{ref:L3189},
and at the HERA ep collider \cite{ref:herasearches}.
If direct production is kinematically forbidden,
the cross-sections of processes 
such as ${\rm e^+e^-} \rightarrow \gamma\gamma$ and
${\rm e^+e^-} \rightarrow {\rm f \bar{f}}$  \cite{ref:opalgg,ref:2f}
are sensitive to new particles at higher masses.

This paper concentrates on the search 
conducted by OPAL \cite{ref:OPAL-detector}
in a wide range of topologies for the pair-production of
new unstable heavy leptons and for both pair- and single-production
of excited leptons of the known generations,
using data collected in 1997
at $\sqrt{s}=181-184$~GeV, with an average energy of 182.7 GeV.
The integrated luminosity used
depends on the final-state topologies,
and is between 52 and 58~pb$^{-1}$.
The results are combined with those obtained earlier from
10~pb$^{-1}$ of data at $\sqrt{s}=$~161.3~GeV
and 10~pb$^{-1}$ at $\sqrt{s}=$~172.1~GeV.

%----------------------------------------------------------------------------

\subsection{Heavy Leptons}
\label{ss:hlintro}

Heavy neutral
leptons are particularly interesting in the light of recent evidence for
massive SM neutrinos~\cite{ref:superk,ref:macro}.  One method of
generating neutrino mass is the see-saw mechanism~\cite{ref:see-saw}, 
which predicts additional heavy neutral leptons.  
In this mechanism, if the mass of a heavy
neutral lepton satisfies the relation 
$m_{\rm N} = m_{\rm e}^2/m_{\nu_{\rm e}}$, 
and if $m_{\nu_{\rm e}}$ is as massive as $2.5$~eV, 
then $m_{\rm N} \approx 100$~GeV, which is within the reach of LEP2.

In general, new heavy leptons N and ${\rm L^\pm}$ could in principle decay
through the charged (CC) or neutral current (NC) channels:
\[
  {\rm N \rightarrow \ell^\pm W^\mp\quad\quad,\quad\quad
       N \rightarrow L^\pm W^\mp\quad\quad,\quad\quad
       N \rightarrow \nu_\ell Z},
\]
\[
  {\rm L^\pm \rightarrow \nu_\ell W^\pm \quad\quad ,\quad\quad
       L^\pm \rightarrow \nul W^\pm \quad\quad,\quad\quad
       L^\pm \rightarrow \ell^\pm Z},
\]
where $\rm \nul$ is a stable or long-lived neutral heavy lepton,
and $\, {\rm \ell = e,\ \mu,\ or\ \tau}$.
For heavy lepton masses less than the gauge boson masses, 
$M_{\rm W}$ or $M_{\rm Z}$, 
the vector bosons are virtual, leading to 3-body decay topologies.
For masses greater than $M_{\rm W}$ or $M_{\rm Z}$, the decays are 2-body decays, and
the CC and NC branching ratios can be comparable.
For masses close to $M_{\rm W}$ or $M_{\rm Z}$, it is important to treat the
transition from the 3-body to the 2-body decay properly, including
effects from the vector boson widths.
Expressions for the computation of partial decay widths with an off-shell
W or Z boson can be found in~\cite{ref:abdel}.

The mixing of a heavy lepton with the standard lepton flavour is governed by
a mixing angle $\zeta$. A mixing of 0.01 radians yields a decay length 
$c\tau$ of $\cal{O}$(1 nm). Since the decay length is proportional 
to $1/\zeta^2$, looking for unstable heavy leptons
which decay within the first cm,
the analyses described in this paper
are sensitive to $\zeta^2 > \cal{O}$(10$^{-12}$).
The presently existing upper limit on
$\zeta^2$ is approximately 0.005 radians$^2$~\cite{ref:nardi}.

The searches for heavy leptons presented in this paper
utilise only the case where N and ${\rm L^\pm}$ decay via the CC channel, 
as would be expected in a naive fourth
generation extension to the SM.  
The NC channel does not contribute significantly
in the heavy lepton searches due to kinematics.
Searches for stable or long-lived charged heavy leptons,
${\rm L^\pm}$,
are described in a separate paper~\cite{ref:stableOPAL183}.

%----------------------------------------------------------------------------

\subsection{Excited Leptons}
\label{ss:elintro}
Compositeness models~\cite{ref:bdk} attempt to explain the
hierarchy of masses in the SM by the existence of a substructure 
within the fermions. Several of these models predict excited 
states of the known leptons.  Excited
leptons are assumed to have the same electroweak SU(2) and U(1)
gauge couplings, $g$ and $g^\prime$, to the vector bosons, 
but are expected to be grouped into
both left- and right-handed weak isodoublets with vector couplings.  
The existence of the right-handed doublets is required to protect the
ordinary light leptons from radiatively acquiring a large
anomalous magnetic moment via the $\ell^*\ell V$
interaction \cite{ref:bdk}
(where $V$ is a $\gamma$, Z, or W$^\pm$ and $\lstar$
refers in this case to both charged and neutral excited leptons).

In ${\rm e^+e^-}$ collisions, excited leptons could be produced in pairs 
via the process ${\rm e^+e^-} \rightarrow \ell^* \bar{\ell}^*$,
or singly via the process ${\rm e^+e^-} \rightarrow \ell^* \bar{\ell}$,
as a result of the $\ell^*\ell V$ couplings.
Depending on the details of these couplings, excited leptons could be
detected in the photonic, CC, or NC channels:
\[
  {\rm \nu_\ell^*  \rightarrow \nu_\ell\gamma     \quad\quad , \quad\quad
       \nu_\ell^*  \rightarrow \ell^\pm W^\mp     \quad\quad , \quad\quad
       \nu_\ell^*  \rightarrow \nu_\ell Z,
  }
\]
\[
  {\rm \ell^{*\pm} \rightarrow \ell^\pm \gamma    \quad\quad , \quad\quad
       \ell^{*\pm} \rightarrow \nu_\ell  W^\pm     \quad\quad , \quad\quad
       \ell^{*\pm} \rightarrow \ell^\pm  Z,
  }
\]
where $\nu^*_{\ell}$ and $\ell^{*\pm}$ are neutral and charged
excited leptons, respectively.

The branching fractions of the excited leptons into the different
vector bosons are determined by the strength of the three
$\ell^*\ell V$ couplings.
We use the effective Lagrangian \cite{ref:bdk}:
\begin{equation}
  {\cal L}_{\ell\ell^*} =
  \frac{1}{2 \Lambda} \bar{\ell}^*\sigma^{\mu\nu}
  \left[g f \frac{ \mbox{\boldmath $\tau$} }{2}
    \mbox{\boldmath {\rm\bf W}}_{\mu\nu} +
  g^\prime f^\prime \frac{Y}{2} B_{\mu\nu} \right] \ell_{\rm L} +
  {\rm hermitian~conjugate},
  \label{eqLll}
\end{equation}
which describes the generalized magnetic de-excitation of
the excited states.
The matrix
$\sigma^{\mu\nu}$ is the covariant bilinear tensor,
\mbox{\boldmath $\tau$} are the Pauli matrices,
${\rm\bf W}_{\mu\nu}$ 
and $B_{\mu\nu}$ represent the fully gauge-invariant
field tensors,
and $Y$ is the weak hypercharge.
The parameter $\Lambda$ has units of energy and can be regarded as
the compositeness scale, while $f$ and $\fp$ are the weights
associated with the different gauge groups.

The relative values of $f$ and $\fp$ also affect the size of the
single-production cross-sections and their detection efficiencies.
Depending on their relative values, either the photonic decay, 
the CC decay, or the NC decay will have the largest
branching fraction, depending on the respective couplings~\cite{ref:bdk}:
\[
 {f_{\gamma} = e_f \fp + I_{3L}(f-\fp)         \quad\quad , \quad\quad
 f_W        = \frac{f}{\sqrt{2} s_w}          \quad\quad , \quad\quad
 f_Z        = \frac{4I_{3L}(c^2_w f + s^2_w \fp) - 4e_fs^2_w \fp}{4 s_w c_w}
}
\]
where $e_f$ is the excited fermion charge, $I_{3L}$ is the weak isospin,
and $s_w(c_w)$ are the sine (cosine) of the Weinberg angle $\theta_w$.

Our results will be interpreted using the two complementary
coupling assignments, $f=\fp$ and $f=-\fp$. For example, 
for the case $f=-\fp$, 
the photonic coupling to excited electrons is suppressed
and the dominant production of excited electrons is via
the $s$-channel. For $f \ne -\fp$, 
the $t$-channel production of excited electrons dominates.
In the case of excited neutrinos, if $f \ne \fp$,
then the photonic coupling is allowed.
In addition to the results for the
two assignments $f=\fp$ and $f=-\fp$,
a new method is introduced which gives limits
on excited leptons independent of the relative values of $f$ and $\fp$.

%============================================================================
\section{Monte Carlo Simulation}
\label{s:mc}

The Monte Carlo (MC) generator EXOTIC~\cite{ref:EXOTIC}
has been used for the simulation of
heavy lepton pair-production, ${\rm e^+e^-}\rightarrow {\rm N \bar N}$ and
${\rm e^+e^-} \rightarrow {\rm L^+L^-}$,
of excited lepton pair-production, 
${\rm e^+e^-}\rightarrow {\rm \nu^*_{\ell} \bar \nu^*_{\ell}}$ and
${\rm e^+e^-} \rightarrow {\rm \ell^{*+}\ell^{*-}}$,
and of single excited lepton production,
${\rm e^+e^-}\rightarrow {\rm \nu^*_{\ell} \bar \nu_{\ell}}$ and
${\rm e^+e^-} \rightarrow {\rm \ell^* \ell}$.
The code is based
on formulae given in~\cite{ref:abdel,ref:zerwas}.
The matrix elements
include all spin correlations in the production and decay processes,
and describe the transition from 3-body to 2-body decays of heavy fermions, 
involving virtual or real vector bosons, including
the effects from vector boson widths.
The JETSET~\cite{ref:jetset} package is used for the fragmentation
and hadronization of quarks. 

For ${\rm N \bar N}$  and ${\rm L^+L^-}$
production, MC samples 
were generated for a set of masses from 40 to 90 GeV.
Separate samples were generated for Dirac and Majorana heavy
neutral leptons, taking into account the different
angular distributions. For the case where ${\rm L^- \rightarrow \nul W^-}$,
samples were simulated at 25 points in the ($M_{\rm L}$,$M_{\nul}$) plane
with $M_{\rm L}$ ranging from 50 to 90 GeV and $M_{\nul}$ from 40 to 87 GeV,
and with a mass difference $M_{\rm L} - M_{\nul}$ larger than 3 GeV.
Excited lepton MC samples were generated
for the pair-production channels with masses in the range from 40 to 90~GeV
and for the single-production channels 
with masses in the range from 90 to 180~GeV.

A variety of MC generators was used to study the multihadronic
background from SM processes:
4-fermion background processes were simulated
using the generator grc4f \cite{ref:grc4f},
multihadronic background from 2-fermion final states was modelled using
PYTHIA \cite{ref:jetset}, while
2-photon processes were generated with PHOJET \cite{ref:phojet}
and  HERWIG \cite{ref:herwig}.
To study the background from low-multiplicity events,
the generators BHWIDE \cite{ref:bhwide} (large-angle Bhabha scattering)
and TEEGG \cite{ref:teegg} ($t$-channel Bhabha scattering)
were used for the $\rm e^+e^- \gamma (\gamma)$ topology. 
The KORALZ generator \cite{ref:koralz}  was used for the
$\mu^+\mu^-\gamma(\gamma)$ and
$\tau^+\tau^-\gamma(\gamma)$ topologies.
These generators include initial and final-state
radiation, which is particularly important for the analyses with
photons in the final states.
Low-multiplicity 4-fermion final states produced in 2-photon interactions
were modelled by using VERMASEREN \cite{ref:vermaseren}.

Finally, SM processes with only photons
in the final state are an important background to the analysis of
excited neutral leptons with photonic decays.
The RADCOR \cite{ref:radcor} program was used to simulate the process
${\rm e^+e^-} \rightarrow \gamma \gamma (\gamma)$ and KORALZ 
was used to simulate the process
${\rm e^+e^-} \rightarrow \nu\overline{\nu} \gamma (\gamma)$. 

All signal and background MC samples were
processed through the full OPAL detector simulation \cite{ref:gopal}
and passed through the same analysis chain as the data.

%============================================================================
\section{Selection}
\label{s:selection}

The searches presented in this paper involve many different
experimental topologies. Three classes of different analyses are 
used which rely on slightly different criteria for
such details as track and cluster quality requirements and 
lepton identification methods. 
The first class of analyses 
includes the selections for high-multiplicity topologies,
with hadronic jets in the final state
from the hadronic CC decays of heavy leptons 
and the hadronic CC and NC decays of excited 
neutral and charged leptons.
The second class includes
selections for low-multiplicity topologies,
and covers the photonic decays of excited charged leptons.
The third class
covers purely photonic event topologies arising from the
photonic decays of excited neutral leptons.

\subsection{High-Multiplicity Topologies}
\label{s:highm}
The searches for topologies with hadronic 
jets in the final state share a common 
high-multiplicity preselection. All charged tracks and calorimeter clusters 
are subjected to established quality criteria \cite{ref:hlOPAL172}. 
Events are required to have at least 8 tracks and 15 clusters.
The total visible energy measured in the detector,
$E_{\rm vis}$, calculated from tracks and clusters~\cite{ref:gce},
must be greater than 20~GeV.

\begin{figure}
\scalebox{0.9}{\includegraphics{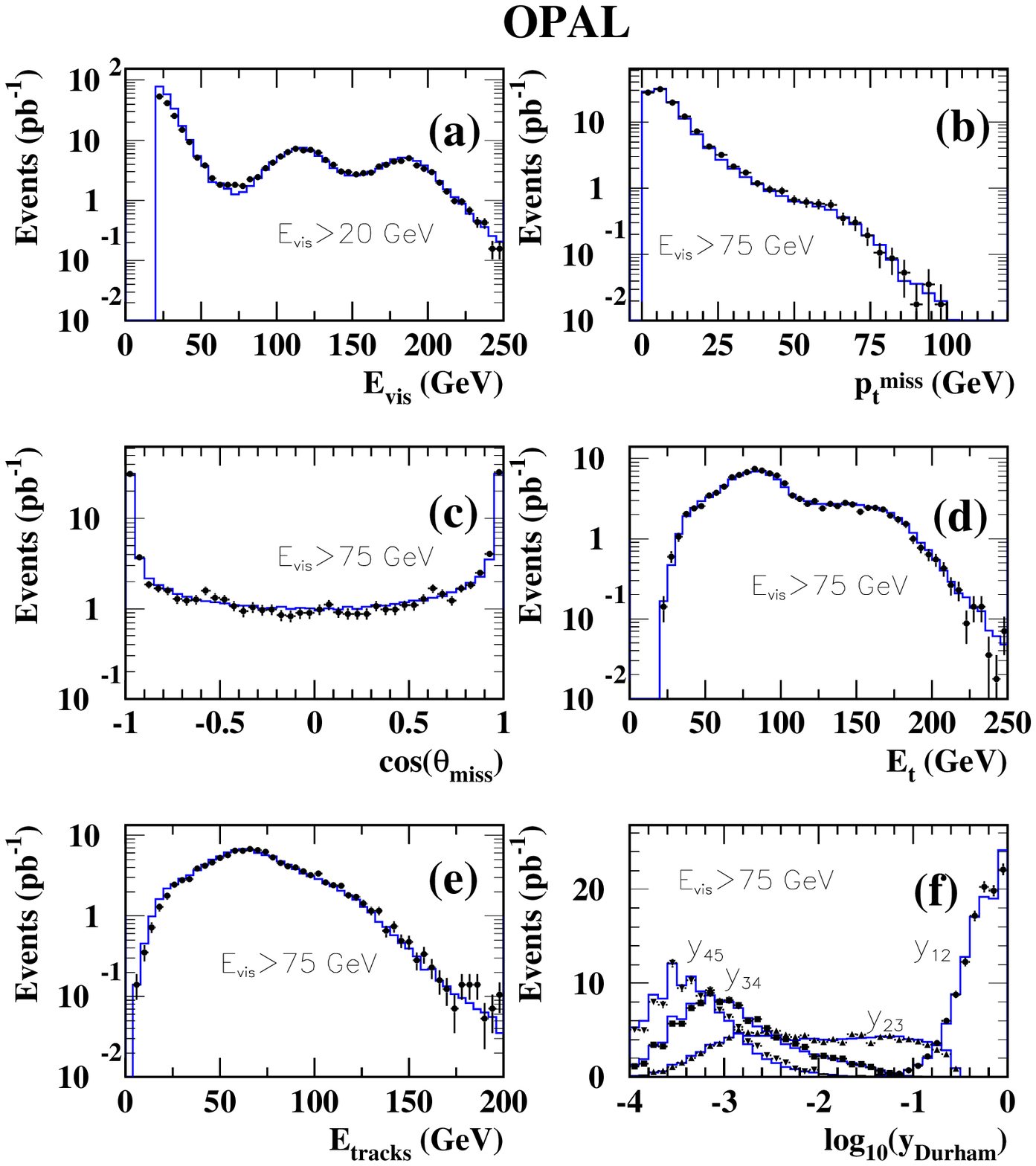}}
\caption{
  Global event properties after preselection cuts for the high-multiplicity
  selections.
  Figure (a) is the visible energy, (b) the missing transverse momentum,
  (c) the cosine of the polar angle of the missing momentum vector, 
  (d) the transverse energy, (e) the summed energy measured from tracks,
  and (f) the logarithm of the jet resolution parameters 
  $y_{12}$, $y_{23}$, $y_{34}$, and $y_{45}$,
  using the Durham jet-finding algorithm~\cite{ref:durham}.
  For figures (b-f) the two-photon background is reduced by requiring
  $E_{\rm vis}>75$~GeV.
  In all figures, the filled symbols are the data and the solid line is
  the SM background prediction. 
}
\label{f:global}
\end{figure}

Global event properties after this preselection
are shown in Figure~\ref{f:global}, which compares data and MC
distributions. 
The visible energy is well described for $E_{\rm vis}>75$~GeV,
where 2-fermion and 4-fermion processes dominate. 
For $E_{\rm vis}<20$~GeV, the data are not as well modelled.
This region is dominated by
events from 2-photon processes, which are not
a significant background to the majority of analyses described in this paper.
After a cut of $E_{\rm vis}>75$~GeV,
the global event properties shown in Figure~\ref{f:global} (b-f) compare well 
between data and the SM expectation.

\begin{figure}
\scalebox{0.9}{\includegraphics{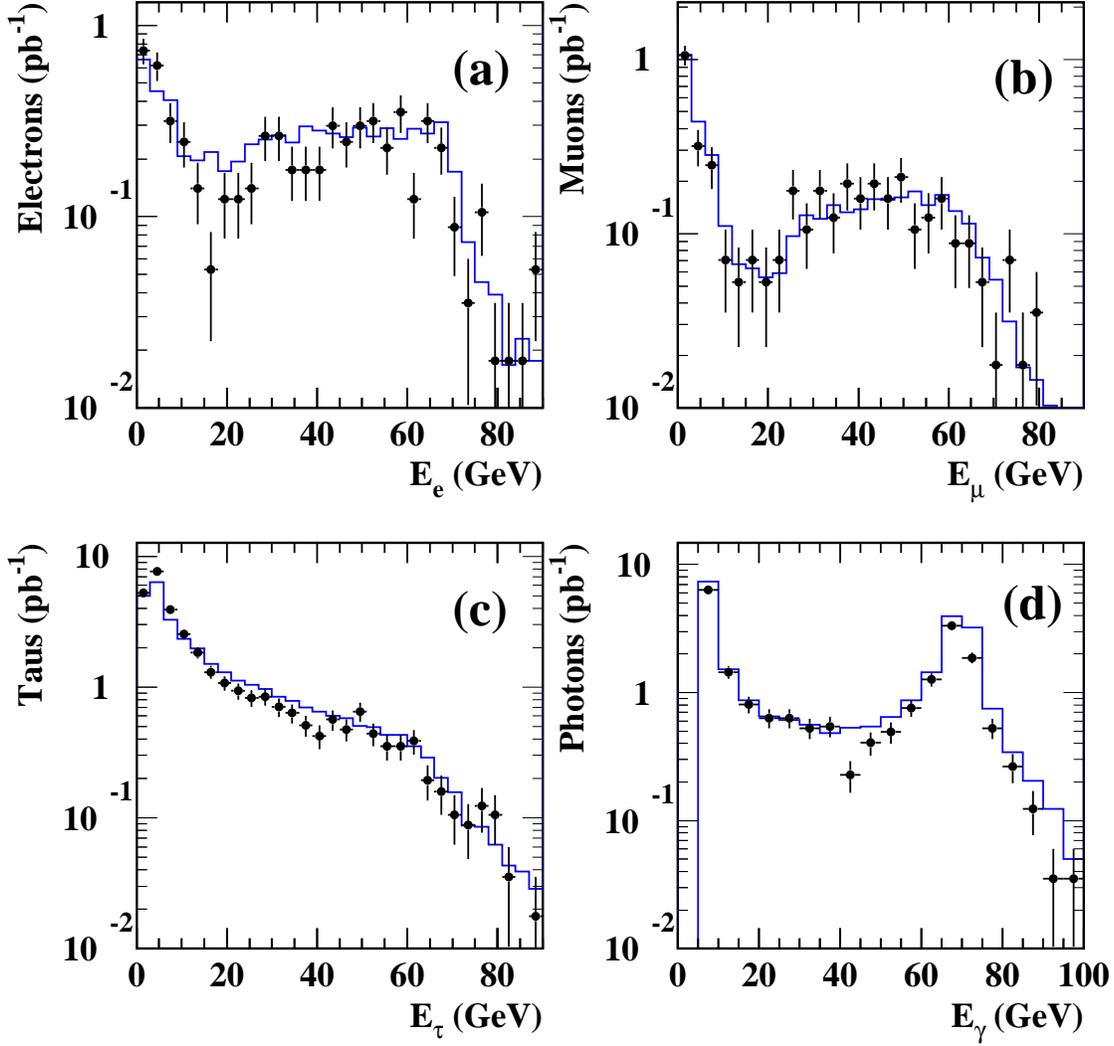}}
\caption{
  Particle identification after preselection cuts.
  The figures show the measured energy of (a) identified electrons,
  (b) identified muons, (c) identified taus, and
  (d) identified photons.
  In all figures, the filled circles are the data and the solid line is
  the sum of all SM background MC.
}
\label{f:lepton}
\end{figure}

Figure~\ref{f:lepton} shows the energy distribution of identified
electrons, muons, taus, and photons after the preselection. 
The lepton identification
is similar to that described in~\cite{ref:hlOPAL172}, with modified
isolation requirements for the high-multiplicity selections.
Identified electrons and muons must have more than 1.5~GeV of visible
energy, and taus more than 3~GeV. In addition,
leptons have to be isolated within a cone of $15^\circ$. 
The typical lepton identification efficiencies are 85-90\% for electrons 
and muons and 70\% for taus.

\subsubsection{Pair-Production of Heavy Leptons}

\paragraph{$\mathbf{L^+ L^-}$ Candidates}\hskip -0.25 cm give rise to final states produced
from flavour-mixing decays into light leptons via ${\rm L \rightarrow \nu_{\rm e} W}$, 
${\rm L \rightarrow \nu_{\rm \mu} W}$, 
and ${\rm L \rightarrow \nu_{\rm \tau} W}$.
The resulting topologies, $\rm \nu \nu W W$, consist of the decay products of
the two W-bosons along with missing transverse momentum.
The main background to this selection is SM W-pair production. 
To optimize the sensitivity, selections for all high-multiplicity final states
$\rm \nu \nu jjjj$, 
$\rm \nu \nu e\nu jj$, 
and $\rm \nu \nu \mu \nu jj$
are performed, where ``j'' refers to a hadronic jet.
The final state
$\rm \nu \nu \tau \nu jj$ does not improve the sensitivity,
and is not considered.
In the selection for $\rm \nu \nu jjjj$ events,
at least 12 GeV of missing momentum and at least 8 GeV of
missing transverse momentum are required, and 
the missing momentum vector must not point along the beam direction
($|{\rm cos(\theta_{miss})}|<0.9$).
In addition, the events must satisfy  higher track (10) 
and cluster multiplicity (35) requirements, 
and are vetoed if a charged lepton is identified.
The number of events after applying these selections is shown in 
Table~\ref{t:pair} for data and for the SM expectation.
The selection efficiency, including the W branching ratio, is about 17-25\%,
depending on the heavy lepton mass.
In the selection of $\nu \nu \rm e \nu jj$ and 
$\nu \nu \rm \mu \nu jj$ 
events, an isolated lepton is required together with
significant visible energy and missing transverse momentum.
In this case, the selection efficiency, including the W branching ratio, is 
about 9-11\%.

\paragraph{$\mathbf{N\bar{N}}$ Candidates}\hskip -0.25 cm  give rise to
final states produced through the
flavour-mixing decay into a light charged lepton,
via ${\rm N \rightarrow e W}$, ${\rm N \rightarrow \mu W}$,
or ${\rm N\rightarrow \tau W}$. 
The topologies are defined as $\rm \ell \ell W W$,
where at least one W-boson decays hadronically and produces jets
in the final state. 
At least two charged leptons of the same flavour are required and
the jet resolution parameters have to be consistent with at least
a 5-jet topology (isolated leptons are treated as ``jets'').
In order to optimize the sensitivity for the case ${\rm N\rightarrow \tau W}$,
the selection has been divided into ${\rm \tau \tau jjjj}$
topologies with fully-hadronic W-decays (2 charged leptons)
and ${\rm \tau \tau \ell \nu jj}$ topologies with semileptonic W-decays,
(3 charged leptons).
The number of events after applying these selections is shown in 
Table~\ref{t:pair}.
The signal efficiencies for a Dirac or Majorana lepton
are about 50\% for ${\rm N \rightarrow e W}$,
57\% for ${\rm N \rightarrow \mu W}$,
and 30-42\% for ${\rm N\rightarrow \tau W}$.

\subsubsection{Pair-Production of Long-Lived Heavy Neutral Leptons}
\label{sss:llNN}
\paragraph{$\mathbf{N_{\rm L} \bar{N}_{\rm L} WW}$ Candidates}\hskip -0.25 cm
originate from the process
${\rm e^+e^-}\rightarrow{\rm L^+L^-}$ with ${\rm L^- \rightarrow \nul W^-}$,
where ${\rm \nul}$ 
is a stable or long-lived neutral heavy lepton which decays 
outside the detector. 
This production is possible if ${\rm \nul}$ is a member of
a fourth-generation SU(2) doublet which does not mix 
with the three known lepton generations and
satisfies $M_{\rm L^\pm} > M_{\rm \nul}$.
This signal leads to very low visible energy if the mass difference,
$\Delta M \equiv M_{\rm L^\pm} - M_{\rm \nul}$, is small.
Events are required to have a visible energy between 8 and 90 GeV,
the missing momentum vector must be at least 20\% of the visible energy
and lie within the barrel region,
a minimum transverse energy of 12 GeV is required,
and the topology should correspond to a pair of acoplanar jets ($>14^{\circ}$).
The total
number of candidates and expected background are shown in Table~\ref{t:pair}.
The typical signal efficiencies are about 30-40\% for $\Delta M>10$~GeV, 
dropping to a few per-cent for $\Delta M=5$~GeV.
A total of 78 events is observed, compared to an expected background of 
52.7 events.  
The discrepancy may be due to mismodelling of the
background, which is dominated by 2-photon processes
with an estimated systematic error on the background of 20\%.
These events typically have energy
deposited in the forward region ($|{\rm cos(\theta)}|>0.9$),
while from the signal MC one does not expect significant forward energy.

\subsubsection{Pair-Production of Excited Leptons with hadronic decays}
Excited leptons, which could be pair-produced at LEP2,
are expected to decay dominantly via CC or photonic interactions,
depending on their coupling assignments.
The final states with both de-excitations via CC decays,
$\rm \ell^* \rightarrow \nu W$ and $\rm \nu^* \rightarrow \ell W$,
are similar to the decay topologies of heavy leptons,
$\rm L \rightarrow \nu W$ and $\rm N \rightarrow \ell W$, 
so the same selections are applied, with
the results shown in Table~\ref{t:pair}.
The doubly-photonic decays give low-multiplicity topologies, and are
discussed in Section~\ref{ss:llgamma}.

\begin{table}
  \footnotesize
\begin{center}
\begin{tabular}{|c|r|r||r|r|} \cline{2-5}
\multicolumn{1}{c|}{} &  Mode & Topology  &  Data  & Total \\
\multicolumn{1}{c|}{} &       &      &        &  Bkd  \\ \cline{2-5} \hline

& $\rm LL \rightarrow \nu\nu WW$ & $\rm \nu \nu W W \rightarrow \nu\nu jjjj$ 
                & 67    &  74.0  \\
& $\rm \ell^*\ell^* \rightarrow \nu\nu WW$ & $\rm \nu \nu W W \rightarrow \nu\nu\nu ejj$
                & 139   & 137.4   \\
CC &  & $\rm \nu \nu W W \rightarrow \nu\nu\nu\mu jj$
              & 111   & 108.1   \\ \cline{2-5}
% &                              & $\rm \nu \nu W W \rightarrow \nu\nu\nu\tau jj$
%                & 240   & 236.0    \\ 
decays 

& $\rm NN \rightarrow \ell\ell WW$ & $\rm NN,\nu^*\nu^* \rightarrow eeWW$
                &  2    &  1.3     \\
& $\rm \nu^*\nu^* \rightarrow \ell\ell WW$ & $\rm NN,\nu^*\nu^*  \rightarrow \mu\mu WW$
                &  3    &  2.4     \\
&                                  & $\rm NN,\nu^*\nu^*  \rightarrow \tau\tau jjjj$
                &  22    &  18.0     \\
&                                  & $\rm NN,\nu^*\nu^*  \rightarrow \tau\tau\nu\ell jj$
                &  2    &  1.2     \\  \cline{2-5}

& $\rm LL \rightarrow \nul\nul WW$ & $\rm \nul \nul W W \rightarrow \nul\nul jjjj$ & 78 & 52.7 \\ \hline

& $\rm \ell^*\ell^* \rightarrow \ell\ell\gamma\gamma$ & $\rm e^* e^*  \rightarrow ee \gamma \gamma$
                &  2    &   3.3    \\
$\gamma$ 
&                 & $\rm \mu^* \mu^*  \rightarrow \mu \mu \gamma \gamma$
                &  1    &   1.1   \\
decays
&                 & $\rm \tau^* \tau^*  \rightarrow \tau \tau \gamma \gamma$
                &  0    &   0.9   \\
& $\rm \nu^*\nu^* \rightarrow \nu\nu\gamma\gamma$ & $\rm \nu^* \nu^*  \rightarrow \nu\nu \gamma \gamma$
                &  8    &   7.4   \\ \hline \hline
\end{tabular}
\end{center}
\caption{Observed number of events in the data sample at $\sqrt{s} =$~183~GeV
and expected
number of events from the background sources in the
searches for the pair-production of heavy and excited leptons,
where ${\rm \nul}$ 
is a stable or long-lived neutral heavy lepton which decays 
outside the detector.}
\label{t:pair}
\end{table}

\subsubsection{Single-Production of Excited Leptons with hadronic decays}
\label{sss:llgamma}
\paragraph{Candidates for the processes $\boldsymbol{\ell}^{\pm *} \boldsymbol{\ell}^{\mp} {\boldsymbol \rightarrow \boldsymbol \ell}^{\pm} {\boldsymbol \ell}^{\mp} {\rm \mathbf Z}$ and ${\boldsymbol{\nu_{\ell}}^{*}} \boldsymbol{ \nu_{\ell} \rightarrow \nu_{\ell} \nu_{\ell} {\rm Z}}$}\hskip -0.25 cm
followed by the hadronic decay of the Z-boson
are selected by requiring two identified leptons of the same flavour
and significant visible energy or 
significant missing transverse momentum ($>$25~GeV), respectively.
In the latter case, events containing charged leptons are vetoed. 
Accepted events have to be consistent with an acoplanar 2-jet topology.
For the charged lepton final states, a kinematic fit is performed
requiring energy and momentum conservation,
and the fit probability has to be consistent with a $\rm \ell^+ \ell^- jj$
final state.
For the $\rm \tau \tau jj$ final state,
to reduce the background further, 
additional cuts on the jet resolution parameter $y_{34}$,
on the missing momentum
vector, and on the ratio of fitted tau energy to visible tau energy ($>1.15$)
are applied.
The results for excited leptons are shown in Table~\ref{t:single} for
data and for the SM expectation.

The selection efficiencies for excited leptons depend on the mass,
and in the case of excited electrons also depend strongly
on the coupling assignments for $f$ and $\fp$ (see Section~\ref{ss:elintro}).
For the case $f=-\fp$,
the $s$-channel production of excited electrons is dominant,
and the selection efficiency is typically 30-50\%.
For the case $f \ne -\fp$,
the $t$-channel production of excited electrons dominates,
in which the scattered electron is preferentially scattered at low angles
and is not detected.
The selection efficiency for this case is typically 7-10\%.
For excited muons and taus only $s$-channel production is allowed and
the selection efficiencies range from 40-50\% and 3-25\%, respectively.
The selection efficiencies for
excited neutrinos are typically 30-40\%.

To improve the sensitivity in the excited electron search for the case
$f \ne -\fp$ a
dedicated selection for the $\rm e(e)jj$ channel has been designed, 
where the scattered electron is not observed in the detector.
Events of this topology are selected by requiring exactly one electron
to be identified.
Tighter isolation cuts on the electron are applied and the event
must have small missing transverse momentum ($< 25$ GeV),
in order to reject $\rm e\nu jj$ final
states from W-pair production.
The event has to be consistent with a 3-jet topology, and
events with high energy photons ($>50$ GeV) in the final state are rejected
to reduce background from radiative returns to the Z.
The fitted kinematics must be consistent with
having an undetected electron in the beam direction
opposite to the detected electron.
The selection efficiency for the case
$f \ne -\fp$ is typically 20-40\% and the numbers
of observed and expected events are shown in Table~\ref{t:single}.

\paragraph{Candidates for the processes $\boldsymbol{\ell}^{\pm} \boldsymbol{\ell}^{\mp*} \boldsymbol{\rightarrow \rm \ell}^{\pm}{\boldsymbol{\nu_{\ell}}}$W$^{\mp}$ and  $\boldsymbol{\nu_{\ell} \nu_{\ell}}^{*} \boldsymbol{\rightarrow \rm \ell}^{\pm} \boldsymbol{\nu _{\ell}}$W$^{\mp}$}
\hskip -0.25 cm
followed by a hadronic decay of the W boson are selected by requiring an
isolated lepton to be identified. The total energy of the event
must be at least 30\% of the centre-of-mass energy.  
For the case $\ell = e$ and $\mu$, the sum of the lepton energy
and the missing transverse momentum must be at least 40\%
of the beam energy.  Background from W-pair production
is reduced by applying a kinematic fit to the ${\rm jj \ell \nu}$ system,
requiring energy and momentum-conservation.
If the resulting masses of the jet-jet and $\ell\nu$ systems 
are consistent with the W mass, the event is rejected.
For the case $\ell = \tau$, 
it is assumed that the
direction of the $\tau$ is given by the direction of the leading particle
of the $\tau$ candidate. Further background suppression is obtained
by requiring that the ratio of energy to mass of the dijet system
be greater than 1.1. The number of observed events and
the SM expectation are shown in Table~\ref{t:single}.
The selection efficiency for these channels, for $f=-\fp$,
is typically from 20--30\%.

For excited electron production in the case $f \ne -\fp$, 
in which
the scattered electron is not observed, a dedicated selection is applied.
In this case, the total energy of the event must be at least 40\% of the
centre-of-mass energy, the missing transverse momentum must be at least
7.5\% of the visible energy,
and there must be no electron identified in the 
event.  The event is forced into two jets, which are required to
be acoplanar ($>50^\circ$). The dijet mass $M_{\rm jj}$
must be consistent with the W mass ($<100$ GeV)
and the ratio of energy to mass of the dijet system must exceed 1.1.
The number of observed events and
the SM expectation are shown in Table~\ref{t:single}.
The selection efficiency for this channel, for $f \ne -\fp$ ,
ranges from 10--30\%.

\subsection{Low-Multiplicity Topologies}
\label{ss:llgamma}
In this section the selection of photonic decays of singly-produced
or pair-produced excited charged leptons is discussed. 
The final states consist of 2 like-flavour leptons and 1 or 2 photons.
The lepton and photon identification and analysis techniques are described
in Reference~\cite{ref:hlOPAL172}.

\subsubsection{Pair-Production of Excited Leptons with photonic decays}

\paragraph{Candidates for the process $\boldsymbol{\ell}^* \boldsymbol{\ell}^* \boldsymbol{\ra \ell}^+ \boldsymbol{\ell}^- \boldsymbol{\gamma \gamma}$}\hskip -0.25 cm
are selected by requiring the identification of two leptons with 
the same flavour and of two photons.  These particles must
carry at least 80\% of the centre-of-mass energy 
in the case of ${\rm e^+e^-}\gamma\gamma$ and $\mu^+\mu^-\gamma\gamma$, 
and between 40\% and 95\% of the centre-of-mass energy 
in the case of $\tau^+\tau^-\gamma\gamma$.
The background in the $\ell^+\ell^- \gamma \gamma$ topology from
Bhabha scattering and di-lepton production
is reduced by requiring the leptons and photons to be 
isolated, and the background from the doubly-radiative return process
$\rm e^+e^- \rightarrow Z \gamma \gamma \rightarrow 
\ell^+ \ell^- \gamma \gamma$
is reduced by vetoing events with di-lepton masses close to the Z mass.
The selection efficiencies are insensitive to the excited lepton mass
and are about 54\% for $\rm e^{*+}e^{*-}$, 61\% for $\rm \mu^{*+}\mu^{*-}$, 
and 40\% for $\rm \tau^{*+}\tau^{*-}$.
The number of observed events and the SM expectation
is shown in Table~\ref{t:pair}.

\subsubsection{Single-Production of Excited Leptons with photonic decays}

\begin{figure}
\scalebox{0.9}{\includegraphics{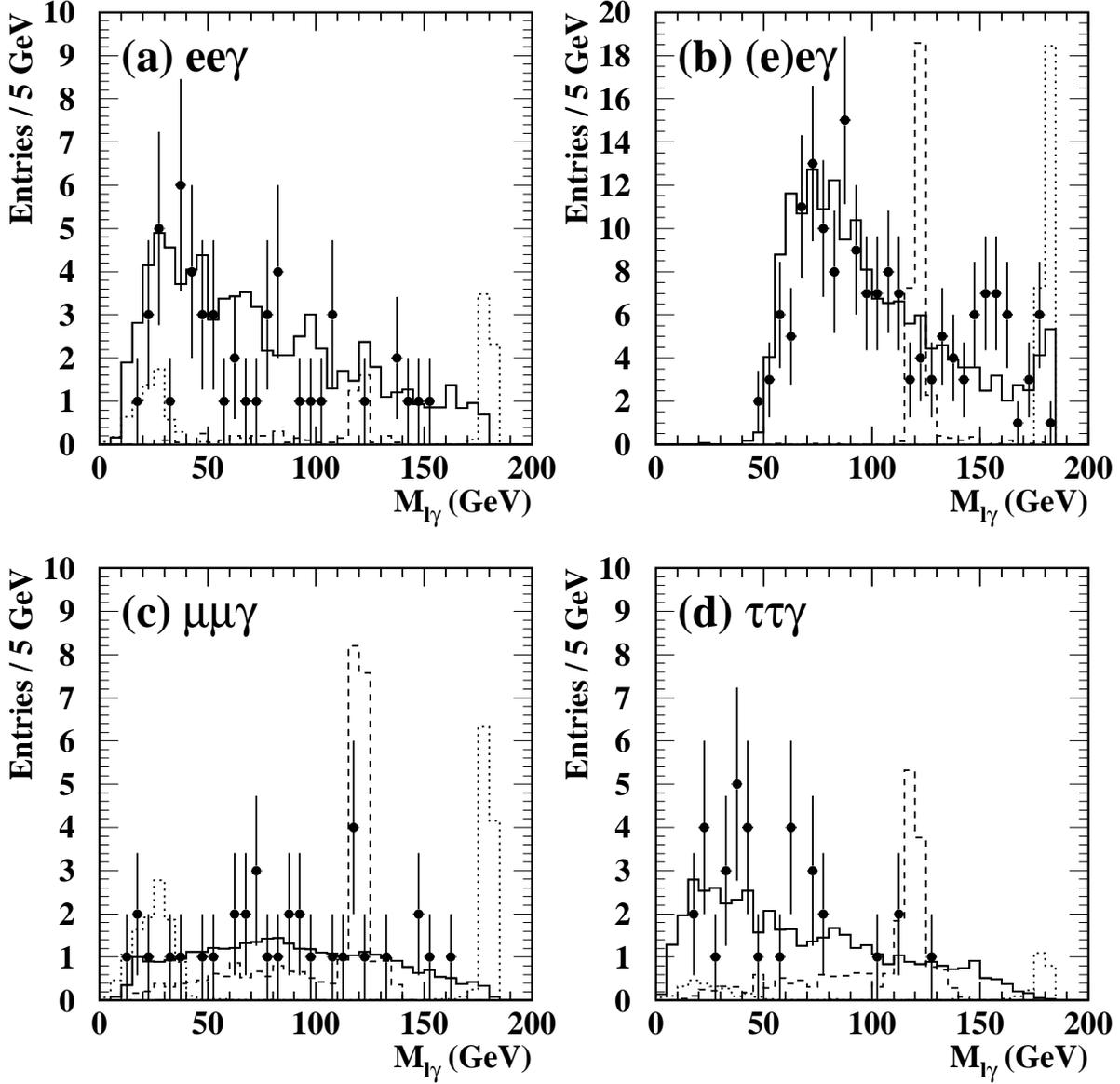}}
\caption{
  Lepton-photon invariant mass distributions for
  (a) ${\rm e^+e^-}\gamma$, (b) ${\rm (e)e}\gamma$,
  (c) $\mu^+\mu^-\gamma$ and (d) $\tau^+\tau^-\gamma$.
  The filled circles are the data, the solid line is
  the sum of all SM background MC, and the
  dashed and dotted lines show example signal MC
  with excited leptons masses of 120 and 180~GeV, respectively.
  There are two entries per event, due to the different mass
  combinations, resulting in a reflection peak at lower
  masses. The signal MC normalization is arbitrary.
}
\label{f:masslg}
\end{figure}

\paragraph{Candidates for the process $\boldsymbol{\ell}^* \boldsymbol{\ell \ra \ell}^+\boldsymbol{\ell}^-\boldsymbol{\gamma}$}\hskip -0.25 cm
are selected with a technique identical
to the $\ell^+\ell^- \gamma \gamma$
selection, except that only one photon is required.  Figure~\ref{f:masslg}
shows the resulting $\ell^{\pm}\gamma$ invariant mass distributions.

To improve the efficiency for the excited electron search
with $f\neq -f^\prime$, giving rise to t-channel production,
a dedicated search is performed for final
states with a single electron and single photon visible in
the detector, assuming the other electron is missing along
the beam axis. Events with one
electron and one photon carrying together at least 40\% of
the total centre-of-mass energy are selected.  The photon is
additionally required to have $|{\rm cos(\theta_{miss})}|<0.7$,
greatly suppressing the Bhabha scattering background.

The number of observed events and the SM expectation
are shown in Table~\ref{t:single}, and the lepton-photon invariant
masses for the selected events are shown in Figure~\ref{f:masslg}.
No peak is observed.  The selection efficiencies 
are typically about 70\% for
${\rm e^*e}$ and $\mu^*\mu$, and about 40\% for $\tau^*\tau$.
The excess in the $\tau^*\tau$ search is consistent with a
statistical fluctuation in the SM $\tau^+\tau^-\gamma$
background.

\subsection{Photonic final states}
\label{ss:phofs}
The search for singly- and pair-produced excited neutral leptons, 
$\rm \nu^* \nu \rightarrow \nu \nu \gamma$ 
and $\rm \nu^* \nu^* \rightarrow \nu \nu \gamma \gamma$,
uses the OPAL search for photonic events with
missing energy, described in~\cite{ref:pr258}.
The numbers of selected events and expected backgrounds are
listed in Tables~\ref{t:pair} and \ref{t:single} for pair- and
single-production, respectively.  For excited neutrinos in
the mass range 70--180~GeV, the selection efficiencies are
70\% and 10--70\% for pair- and single-production, respectively.

\begin{table}
  \footnotesize
\begin{center}
\begin{tabular}{|c|r|r||r|r|} \cline{2-5}
\multicolumn{1}{c|}{} &  Mode & Topology  &  Data  & Total \\
\multicolumn{1}{c|}{} &       &      &        &  Bkd  \\ \cline{2-5} \hline
& $\rm \ell^* \ell \rightarrow \ell \ell Z$ &  ${\rm e^*e \rightarrow e e \, jj}$ 
                &    2 &    2.3 \\ 
{NC} &  & ${\rm e^*e \rightarrow e (e) \, jj}$ 
                &   32 &    27.3 \\ 
{decays} & & ${\rm \mu^*\mu \rightarrow \mu \mu \, jj}$ 
                &    2 &    2.2 \\ 
& & ${\rm \tau^*\tau \rightarrow \tau \tau \, jj}$ 
                &    5 &    4.3  \\ \cline{2-5} 
& $\rm \nu^* \nu \rightarrow \nu \nu Z$ & ${\rm \nu^*\nu \rightarrow \nu \nu \, jj}$ 
                &   23 &    22.9  \\ \hline \hline
& $\rm \ell^* \ell \rightarrow \nu \ell W$ &  ${\rm e^*e \rightarrow \nu_e e \,  jj}$ 
                &    13 &    18.4  \\ 
&  &  ${\rm e^*e \rightarrow \nu_e (e) \, jj}$ 
                &    18 &    17.6  \\ 
& & ${\rm \mu^*\mu \rightarrow \nu_\mu \mu jj}$ 
                &     6 &     6.7  \\
{CC} & & ${\rm \tau^*\tau \rightarrow \nu_\tau \tau jj}$ 
                &    23 &    27.7  \\  \cline{2-5} 
{decays} & $\rm \nu^* \nu \rightarrow \nu \ell W$  & ${\rm \nu_e^*\nu_e \rightarrow \nu_e e \, jj}$ 
                &  13   &   18.4  \\ 
& & ${\rm \nu_\mu^*\nu_\mu \rightarrow \nu_\mu \mu jj}$ 
                &   6   &   6.7   \\ 
& & ${\rm \nu_\tau^*\nu_\tau \rightarrow \nu_\tau \tau jj}$ 
                & 23   &  27.7    \\ \hline \hline
& $\rm \ell^* \ell \rightarrow \ell \ell \gamma$ &  ${\rm e^*e \rightarrow e e \, \gamma}$ 
                &   25 & 39.9 \\ 
{$\gamma$} & & ${\rm e^*e \rightarrow (e) e \, \gamma}$ 
                &   195 &    213.7  \\ 
{decays} & & ${\rm \mu^*\mu \rightarrow \mu \mu \, \gamma}$ 
                &    17 &    16.5 \\ 
& & ${\rm \tau^*\tau \rightarrow \tau \tau \, \gamma}$ 
                &   34 &    22.9  \\ \cline{2-5} 
& $\rm \nu^* \nu \rightarrow \nu \nu \gamma$ & $ {\rm \nu^*\nu \rightarrow \nu \nu \, \gamma}$ 
                &   4 &    3.5  \\ \hline \hline
\end{tabular}
\end{center}
\caption{Observed number of events in the data sample at $\sqrt{s} =$~183~GeV 
and expected number of events from the background sources for the
searches for the single-production of excited leptons.
The symbol (e) indicates topologies in which one scattered electron is
not observed in the detector.}
\label{t:single}
\end{table}

%============================================================================
\section{Results}
\label{s:results}

The numbers of expected signal events are evaluated from the 
production cross-sections, the integrated luminosity, and
the estimated detection efficiencies of the various analyses.

The systematic errors on the number of expected
signal and background events are estimated from: 
     the statistical error of the
     MC estimates (1-10\%),
     the error due to the interpolation used to infer the
     efficiency at arbitrary masses from a limited number
     of MC samples (2-15\%),
     the error on the integrated luminosity (0.6\%),
     the uncertainties in modelling the lepton identification
     cuts and in the photon conversion finder efficiency (2-8\%), 
     and the error due to uncertainties of
     the energy scale, energy resolution, and error parameterisations (1-4\%).
The errors are considered to be independent and are added in
quadrature to give the total systematic error which is taken into account
for the limit calculations.

In the case of pair-production searches, the production cross-section
is relatively model-independent, and limits on the masses of
the heavy or excited leptons can be obtained directly.
For the flavour-mixing heavy lepton decays in which the
pair-produced heavy leptons undergo CC decays into
light leptons, 95\% confidence level (CL) lower limits on the mass
of the heavy lepton are shown in Table~\ref{t:hlmasslim}.
The results are given for both Dirac and Majorana
heavy neutral leptons.
These results are valid for a mixing angle squared, $\zeta^2$,
greater than about 10$^{-12}$ radians$^2$.
For the decays of charged heavy leptons into a massless neutral lepton,
${\rm L\rightarrow \nu_\ell W}$, the searches 
for the $\rm \nu \nu jjjj$ decay 
and for the $\rm \nu \nu jj \ell \nu_{\ell}$ have been combined.
Masses smaller than
84.1~GeV are excluded at the 95\% CL.

\begin{figure}
\scalebox{0.9}{\includegraphics{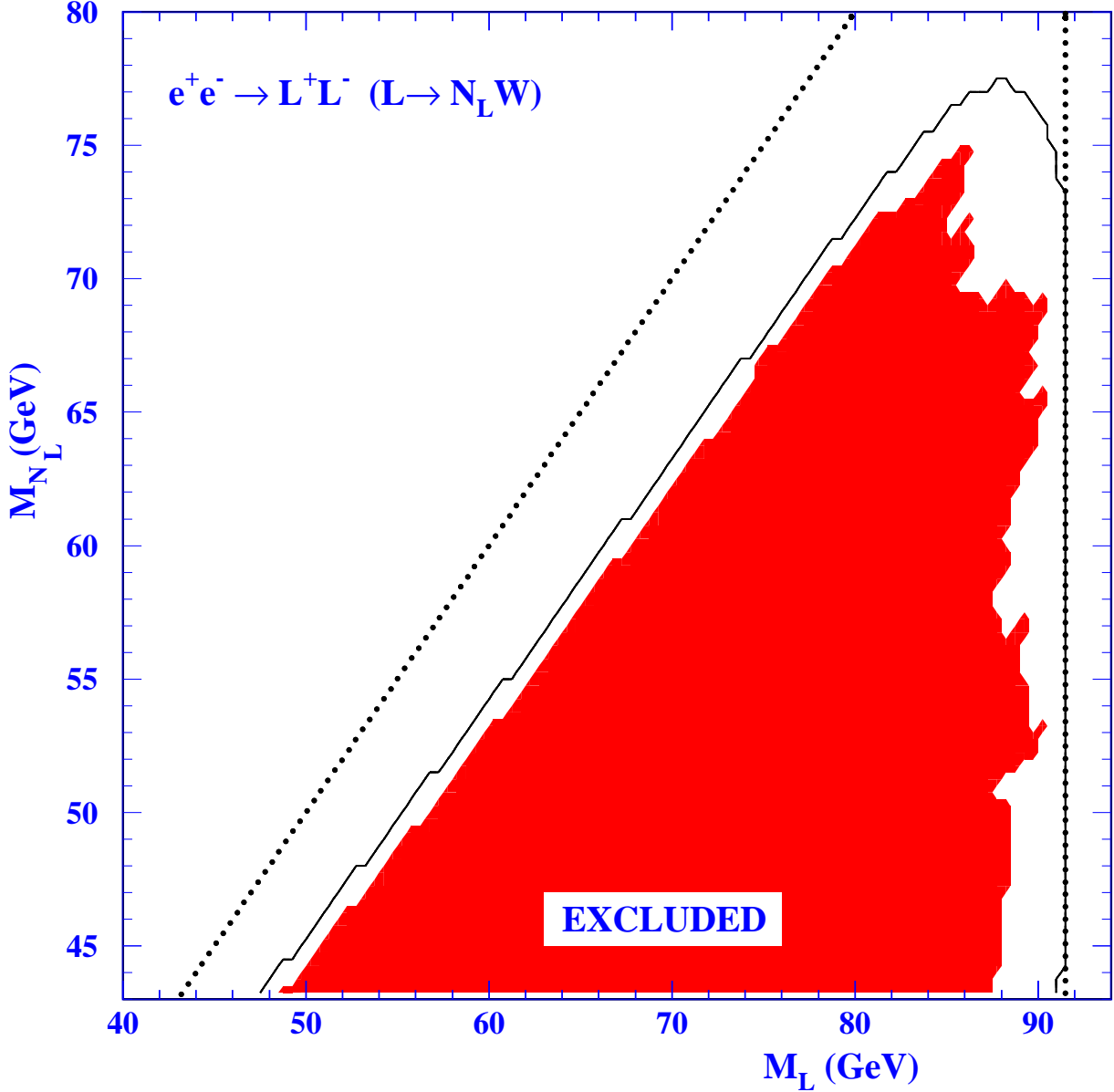}}
\caption{Excluded region in the $(M_{\rm L}, M_{\rm \nul} )$ plane for $\LL$
  production in the ${\rm L^- \rightarrow \nul W^-}$ case,
  where ${\rm \nul}$ 
  is a stable or long-lived neutral heavy lepton which decays 
  outside the detector.  The filled area represents the region excluded 
  at the 95\% CL.
  The dashed vertical line represents the kinematic limit,
  while the diagonal line corresponds to
  $\Delta M \equiv M_{\rm L} -M_{\rm \nul}=0$.}
  The dark contour line shows the mean expected limit.
\label{f:mlmnexcl}
\end{figure}

In the case that a heavy charged lepton ${\rm L^{\pm}}$ decays into
a stable heavy neutral lepton, $\nul$, the exclusion region depends on
both $M_{\rm L}$ and $M_{\rm \nul}$.  In order to optimize
the sensitivity of the analysis, a scan is performed in steps of $\Delta M$. 
At each point, the expected minimum and maximum visible energy,
(${\rm E_{min}, E_{max}}$), are calculated analytically,
and the corresponding number of observed and expected events is determined. 
The resulting region in ($M_{\rm L},M_{\rm N}$) excluded at the 95\% CL
is given in Figure~{\ref{f:mlmnexcl}}, together with the mean expected limit.

\begin{table}
\begin{center}
\begin{tabular}{|c|c|c|}\hline
   \multicolumn{2}{|c|}{Mode}                             & Mass Limit                       \\
    \multicolumn{2}{|c|}{}                                &  (GeV)                           \\ \hline\hline
${\rm N \rightarrow e W}$           & Dirac               &  88.0                            \\
                                    & Majorana            &  76.0                            \\ \hline
${\rm N \rightarrow \mu W}$         & Dirac               &  88.1                           \\
                                    & Majorana            &  76.0                            \\ \hline
${\rm N \rightarrow \tau W}$        & Dirac               &  71.1                            \\
                                    & Majorana            &  53.8                            \\ \hline
\multicolumn{2}{|c|}{${\rm L\rightarrow \nu_\ell W}$} &  84.1                            \\ \hline
%\multicolumn{2}{|c|}{${\rm L^-\rightarrow N W^-}$}        &  81.5 (with $\Delta M > 8.4$)  \\
\hline
\end{tabular}
\caption{95\% CL lower mass limits
  on unstable neutral and charged heavy leptons
  obtained from the data
  collected at $\protect\sqrt{s}=$~183~GeV.}
\label{t:hlmasslim}
\end{center}
\end{table}

The mass limits on excited leptons are somewhat better than
for the heavy lepton case, primarily due to the nature of the vector couplings
which lead to larger production cross-sections~\cite{ref:bdk}.  The mass
limits inferred from the pair-production searches
are shown in Table~\ref{t:lstar_masslim} for charged excited leptons
and in Table~\ref{t:nustar_masslim} for neutral excited leptons.  
The first two sections of each table give the mass limits
in which the dominant decay mode is assumed to be either
via photons or W bosons.

\begin{table}
\begin{center}
\begin{tabular}{|c|c|l|c|} \hline
   Flavour         &   Coupling    & Dominant      & Mass Limit \\
                   &               & Decay         & (GeV)      \\ \hline\hline
   ${\rm e}^*$     & $f=f^\prime$  & Photonic      & 91.3       \\
   $\mu^*$         & $f=f^\prime$  & Photonic      & 91.3       \\
   $\tau^*$        & $f=f^\prime$  & Photonic      & 91.2       \\ \hline
   ${\rm e}^*$     & $f=-f^\prime$ & CC            & 86.0       \\
   $\mu^*$         & $f=-f^\prime$ & CC            & 86.0       \\
   $\tau^*$        & $f=-f^\prime$ & CC            & 86.0       \\ \hline
   ${\rm e}^*$     & \multicolumn{2}{c|}{Any $f$ and $f^\prime$} & 84.6\\
   $\mu^*$         & \multicolumn{2}{c|}{Any $f$ and $f^\prime$} & 84.7\\
   $\tau^*$        & \multicolumn{2}{c|}{Any $f$ and $f^\prime$} & 84.5\\\hline\hline
\end{tabular}
\end{center}
\caption{ 95\% CL lower
  mass limits for the different charged excited leptons obtained from the
  pair production searches.  The coupling assumption
  affects the dominant branching ratio.
  }
\label{t:lstar_masslim}
\end{table}

\begin{table}
\begin{center}
\begin{tabular}{|c|c|l|c|} \hline
   Flavour         &   Coupling    & Dominant      & Mass Limit \\
                   &               & Decay         & (GeV)      \\ \hline\hline
   $\nu_{\rm e}^*$ & $f=f^\prime$  & CC            & 91.1       \\
   $\nu_\mu^*$     & $f=f^\prime$  & CC            & 91.1       \\
   $\nu_\tau^*$    & $f=f^\prime$  & CC            & 83.1       \\ \hline
   $\nu_{\rm e}^*$ & $f=-f^\prime$ & Photonic      & 91.2       \\
   $\nu_\mu^*$     & $f=-f^\prime$ & Photonic      & 91.2       \\
   $\nu_\tau^*$    & $f=-f^\prime$ & Photonic      & 91.2       \\ \hline
   $\nu_{\rm e}^*$ & \multicolumn{2}{c|}{Any $f$ and $f^\prime$} & 90.5  \\
   $\nu_\mu^*$     & \multicolumn{2}{c|}{Any $f$ and $f^\prime$} & 90.5  \\
   $\nu_\tau^*$    & \multicolumn{2}{c|}{Any $f$ and $f^\prime$} & 81.5 \\ \hline\hline
\end{tabular}
\end{center}
\caption{ 95\% CL lower
  mass limits for the different neutral excited leptons obtained from the
  pair production searches.  The coupling assumption affects
  the dominant branching ratio.
  }
\label{t:nustar_masslim}
\end{table}

In the single-production searches, the production cross-section depends
on parameters within the model, so limits on those parameters, as a 
function of the new particle masses, are inferred instead.
From the single production searches, 95\% CL
limits have been calculated on $\sigma \times {\rm BR}$ for
${\rm e^+e^-} \rightarrow \ell^* \bar{\ell}, ~\ell^* \rightarrow \ell V$
for the different particle types, including photonic decay results from  
$\protect\sqrt{s}=$~161~GeV and 172~GeV,
where the branching ratio, ``BR'', depends
on the relative values of $f$ and $\fp$.

\begin{figure}
\scalebox{0.9}{\includegraphics{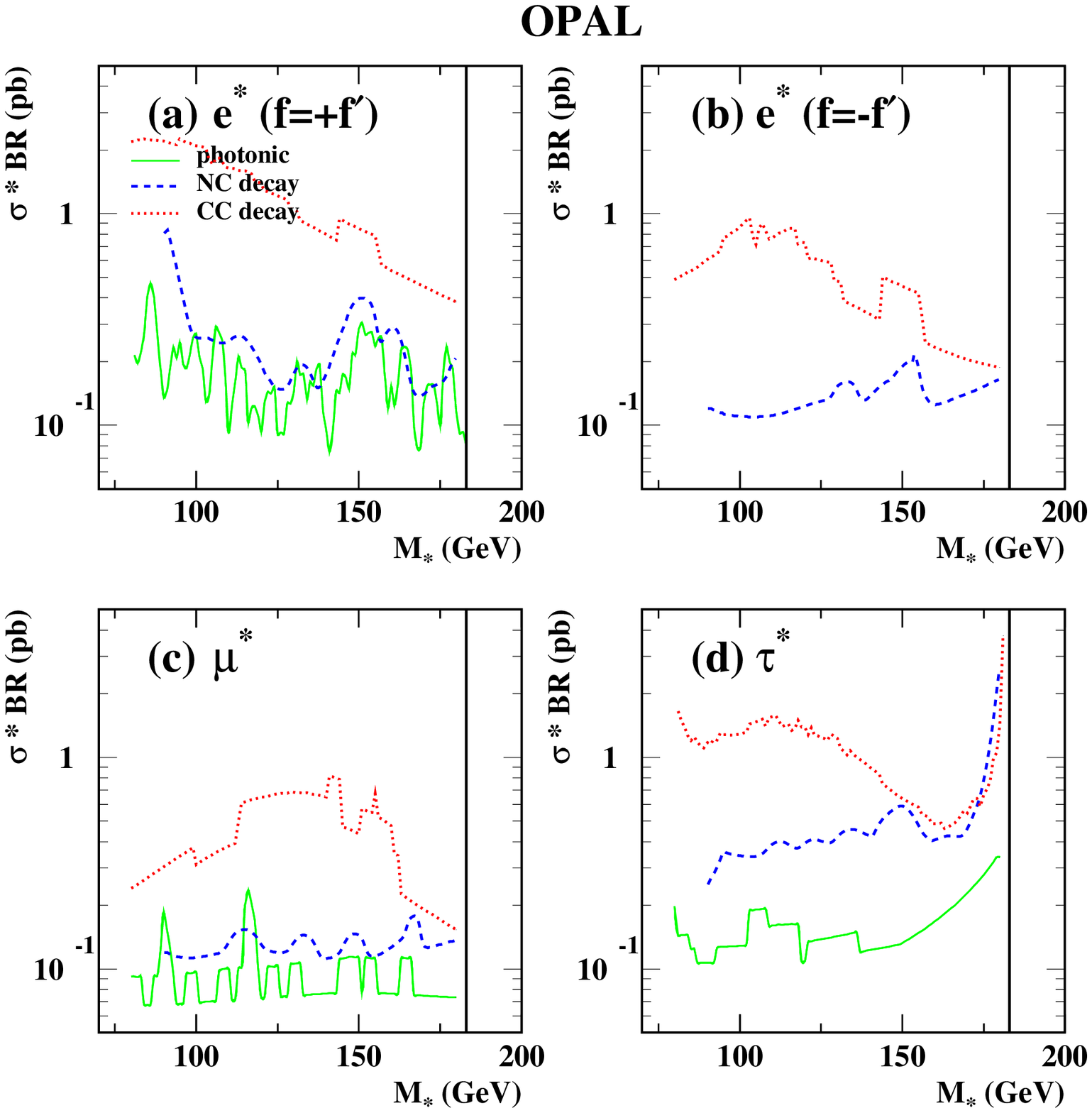}}
\caption{Upper limits at the 95\% CL on the $\sigma \times BR$
  of ${\rm e^+e^-} \rightarrow \ell^* \bar{\ell}, ~\ell^* \rightarrow \ell V$
  for the single-production of 
  (a) $\rm e^*$ with $f=f^\prime$, 
  (b) $\rm e^*$ with $f=-f^\prime$, 
  (c) $\mu^*$, and (d) $\tau^*$.
  The photonic decay is represented by the full line,
  the NC decay by the dashed line,
  and the CC decay by the dotted line.
  For single $\mu^*$ and $\tau^*$ production the selection efficiency
  does not rely on the coupling assignment.
  For $f=-f^\prime$ the photonic decay of $\estar$ is forbidden.
  The symbol ${\rm M_*}$ represents the mass of the excited lepton.
  }
\label{f:sbrlim1}
\end{figure}

\begin{figure}
\scalebox{0.9}{\includegraphics{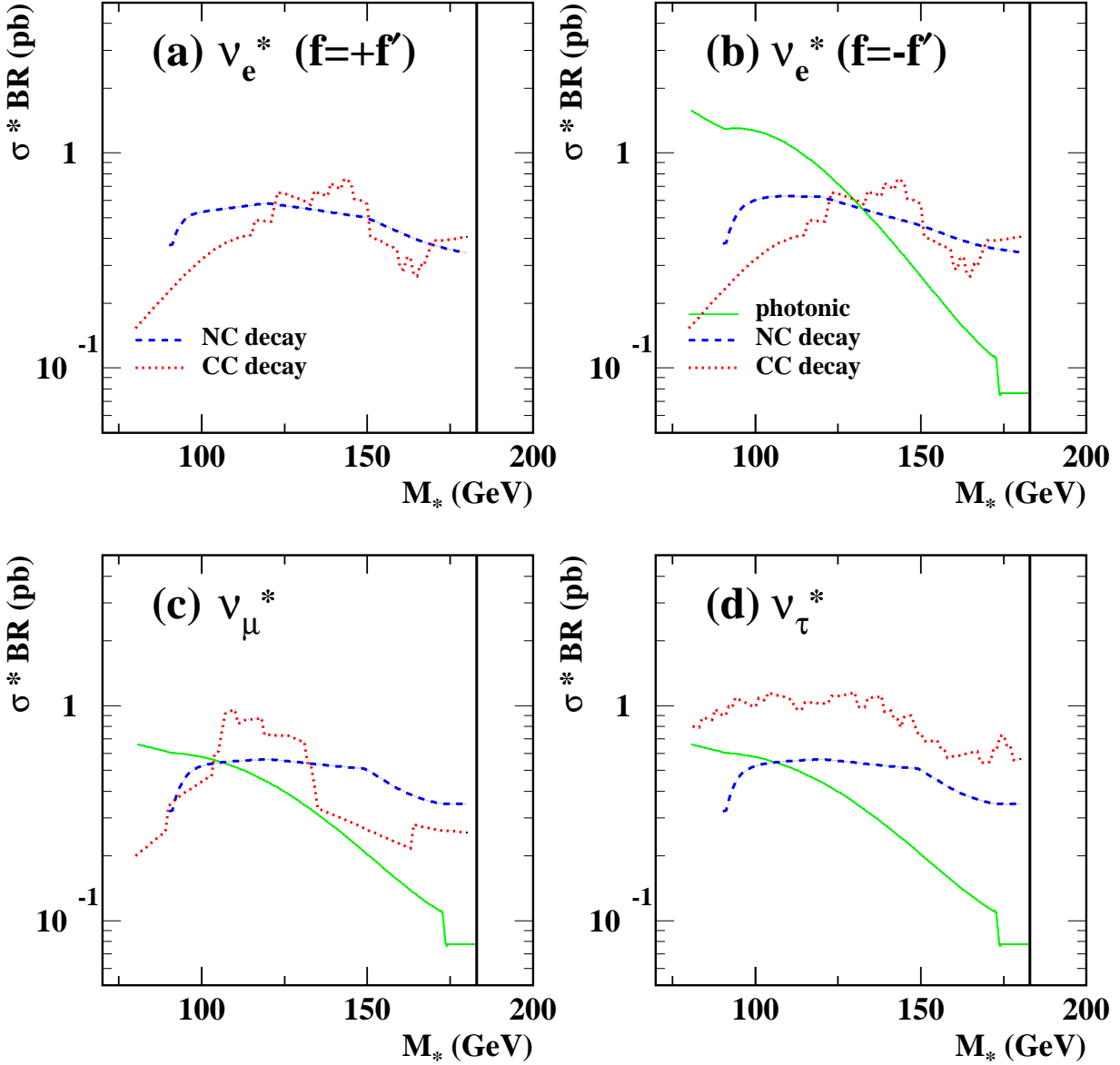}}
\caption{Upper limits at the 95\% CL on the $\sigma \times BR$
  of ${\rm e^+e^-} \rightarrow \ell^* \bar{\ell}, ~\ell^* \rightarrow \ell V$
  for the single-production of 
  (a) $\nu_{\rm e^*}$ with $f=f^\prime$, 
  (b) $\nu_{\rm e^*}$ with $f=-f^\prime$, 
  (c) $\nu_\mu^*$, and (d) $\nu_\tau^*$.
  The photonic decay is represented by the full line,
  the NC decay by the dashed line,
  and the CC decay by the dotted line.
  For single $\nu_\mu^*$ and $\nu_\tau^*$ production the selection efficiency
  does not rely on the coupling assignment.
  For $f=f^\prime$ the photonic decay of $\nu_{\rm e}^*$ is forbidden.
  The symbol ${\rm M_*}$ represents the mass of the excited lepton.
  }
\label{f:sbrlim2}
\end{figure}

In general, the mass limits have been evaluated using a sliding window 
technique~\cite{ref:hlOPAL172}, 
taking into account the expected mass resolution 
for the $\rm e\nu jj$, $\rm \mu\nu jj$, 
$\rm \tau\nu jj$ topologies (CC decays) and 
the $\rm ee\gamma$, $\rm \mu \mu \gamma$, $\rm \tau \tau \gamma$ topologies 
(photonic decays), 
or by taking into account the kinematically allowed mass limits for
the $\nu \nu \gamma$ topologies (photonic decays). 
For the $\rm eejj$, $\rm \mu\mu jj$, and $\rm \tau \tau jj$ topologies 
(NC decays),
a likelihood fit method
has been used for the limit calculation~\cite{ref:lyons}.
The resulting limits on $\sigma \times {\rm BR}$ 
are shown for excited charged leptons in Figure~\ref{f:sbrlim1}
and for excited neutral leptons in Figure~\ref{f:sbrlim2} for all
generations
decaying via photonic, neutral current, and charged current processes.
The $\sigma \times {\rm BR}$ limits do not depend on the coupling
assignments except for excited electrons,
where the selection efficiencies depend on the ratio of $t$-channel
and $s$-channel  contributions,
and the results are shown for the example assignments $f=\pm f'$.
The limits for the photonic decays are valid only for one of the
two coupling assignments,
$f=+f'$ for excited charged leptons and $f=-f'$ for excited neutral
leptons.

\begin{figure}
\scalebox{0.9}{\includegraphics{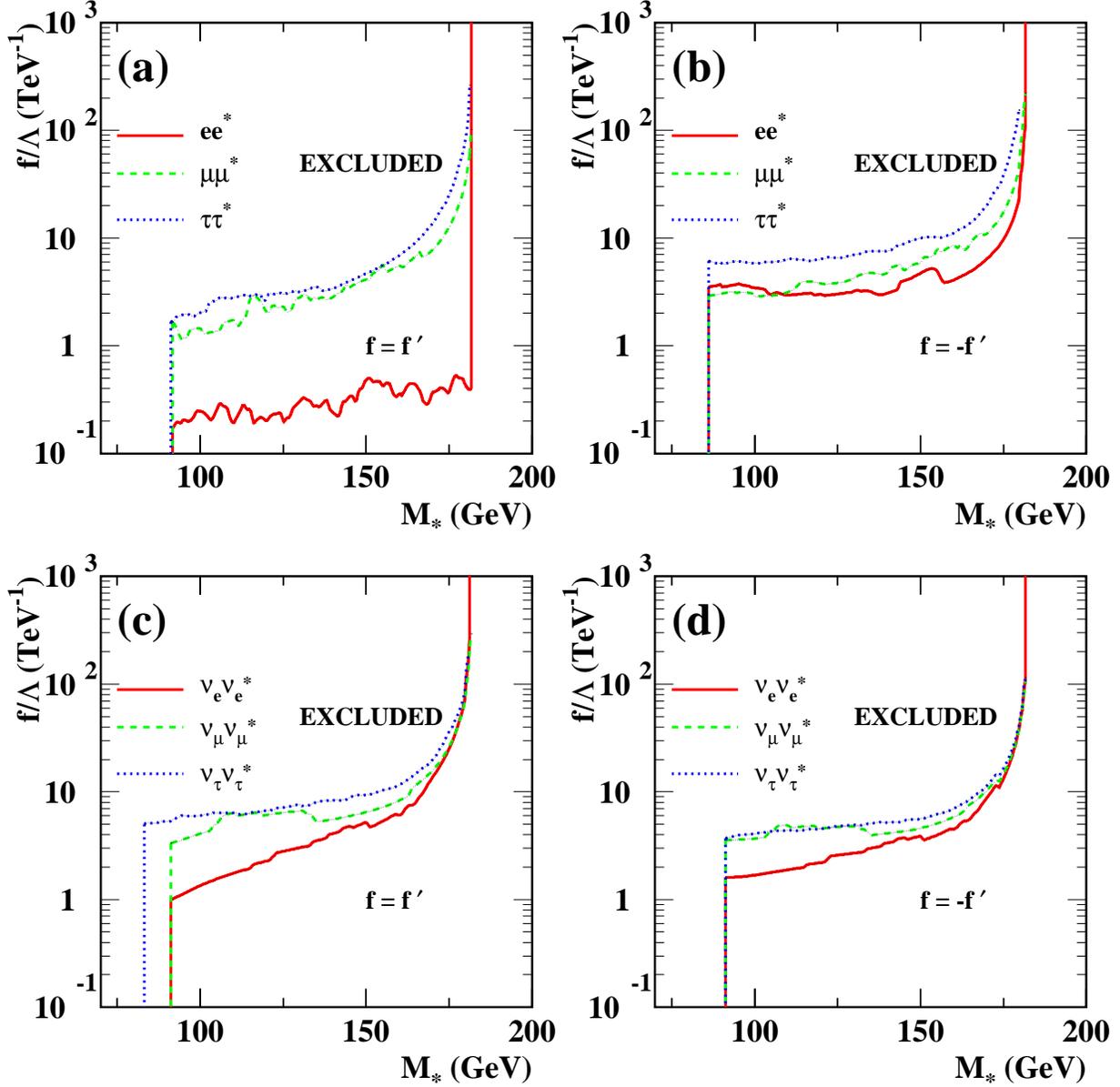}}
\caption{Upper limits at the 95\% CL on the ratio $f/\Lambda$ 
  of the coupling strength to the compositeness scale,
  as a function of the
  excited lepton mass.  (a) The limits on ${\rm e}^*$,
  $\mu^*$ and $\tau^*$ with $f=f^\prime$;
  (b) the limits on  ${\rm e}^*$, $\mu^*$ and $\tau^*$ with $f=-f^\prime$,
  (c) the limits on $\nu_{\rm e}^*$, $\nu_\mu^*$ and $\nu_\tau^*$ with
  $f=f^\prime$,
  and (d) the
  limits on $\nu_{\rm e}^*$, $\nu_\mu^*$ and $\nu_\tau^*$ with
  $f=-f^\prime$.
  The regions above and to the left of the
  curves are excluded by the single- and pair-production searches,
  respectively.
  The symbol ${\rm M_*}$ represents the mass of the excited lepton.
  }
\label{f:elcouplim}
\end{figure}

From the single-production
searches, limits on the ratio of the coupling to the compositeness
scale, $f/\Lambda$, can be inferred.
The results are shown in Figure~\ref{f:elcouplim} for
two coupling assumptions\footnote{
  For the figures of coupling versus compositeness scale 
  given in~\cite{ref:hlOPAL172}, 
  an error was discovered in the cross-section formula
  used; the resulting limits on $f/\Lambda$ were over-conservative for
  most values of the excited lepton mass.  This error 
  has been corrected in the present paper.}.
Since the branching ratio of
the excited lepton decays via the different vector bosons is
not known, examples of coupling assignments, $f=\pm f^\prime$,
are used to calculate these branching ratios and then the
photonic, NC and CC decay results are combined for the limits.

A new method is used to infer limits on the coupling strength
$f_0/\Lambda$, independently of
the relative values of $f$ and $f^\prime$.
Here $f_0$ is a generalized coupling constant defined as
$f_0=\sqrt{ \frac{1}{2}(f^2+{f^\prime}^2)}$. 
It is also useful to define the parameter $\tan \phi_f = f/{f^\prime}$;
the previous coupling assignments
then correspond to $\phi_f=\pi/4$~($f=\fp$)
and $\phi_f=-\pi/4$~($f=-f^\prime$).

From the Lagrangian in Equation~\ref{eqLll},
the cross-section depends on $f$ and $f^\prime$ in the following
way:
\begin{eqnarray}
\sigma 
\  =  \frac{\ a_1 \; f^2 \; + \; a_2 \; f \; f^\prime \; + \; a_3 \; {f^\prime}^2}{\Lambda^2}  =  \bigl(\frac{f_0}{\Lambda}\bigr)^2  \cdot A(\phi_f),
\end{eqnarray}
where $ A(\phi_f)= 2 \cdot (a_1 \sin^2{\phi_f} +
                                     a_2 \sin{\phi_f}\cos{\phi_f} +
                                     a_3 \cos^2{\phi_f} )$.
The coefficients $a_1$, $a_2$, and $a_3$ can be calculated from the matrix elements.
The production cross-section
and also the decay of the excited leptons depend on the
coupling assumption and on the angle $\phi_f$. The branching ratio is
calculated using the formulae given in~\cite{ref:bdk}.
By combining both the production cross-section and the branching ratio,
the likelihood function $L(N_s)$, which is a function of the number of 
observed signal events $N_s$, can be translated into a likelihood function
depending on the coupling strength $\frac{f_0}{\Lambda}$:
\begin{equation}
  L  \bigl( \frac{f_0^2}{\Lambda^2} \bigr) \
  = \ L \bigl( \frac{N_s} {A(\phi_f) \cdot BR(\phi_f) \cdot \epsilon(\phi_f) \cdot {\cal L}} \bigr) \ \ ,
  \label{eq:Logl1}
\end{equation}
where $\epsilon(\phi_f)$ is the total selection efficiency and $\cal L$
the integrated luminosity.

In the case of $\mstar$, $\tstar$, $\nustarm$, and $\nustart$ production
the efficiency is constant, while
the case of the single production of excited $\rm e^*$
and $\nu_{\rm e}^*$ is more complicated. Due to the different
$\phi_f$ - dependent contributions of the
$t$-channel and the $s$-channel diagrams, the selection efficiency also depends
on $\phi_f$. 
Using a MC technique to determine the  selection efficiency for arbitrary
$\phi_f$ would not be practical because a large number of MC events would
have to be simulated for different excited lepton masses.
It turns out, however, that the selection efficiency can be written as:
\begin{eqnarray}
\epsilon(\phi_f) \  =  \ \frac{\sigma_{sel}}{\sigma_{gen}} =
\ \frac{e_1 \; f^2 \; + \; e_2 \; ff^\prime \; + \; e_3 \; {f^\prime}^2}
{a_1 \; f^2 \; + \; a_2 \; ff^\prime \; + \; a_3 \; {f^\prime}^2}
= \frac{N(\phi_f)}{A(\phi_f)},
\end{eqnarray}
with $N(\phi_f)= 2 \cdot (e_1 \sin^2{\phi_f} \; + \; e_2 \sin{\phi_f} \cos{\phi_f} \; + \; e_3 \cos^2{\phi_f} )$.
The coefficients $e_i$ can be calculated by evaluating the
selection efficiencies from three MC samples generated with different
values of $\phi_f$.

The selection efficiencies for the different decay topologies
and the cross-section vary strongly with $\phi_f$
in the case of single $\rm e^*$ production. 
In order to avoid numerical errors, one of the generated MC points 
has been chosen for the coupling assumption $f=-f^\prime$, where
the large $t$-channel contribution vanishes and the cross-section is smallest.
After having calculated the coefficients describing the selection efficiencies,
the error of this method has been tested by comparing
the calculated selection efficiency for a given
value of $\phi_f$ with the selection efficiency determined from a
MC sample generated with the same $\phi_f$ value.
The error is found to be small compared to the statistical error 
of the MC samples.

\begin{figure}
\scalebox{0.9}{\includegraphics{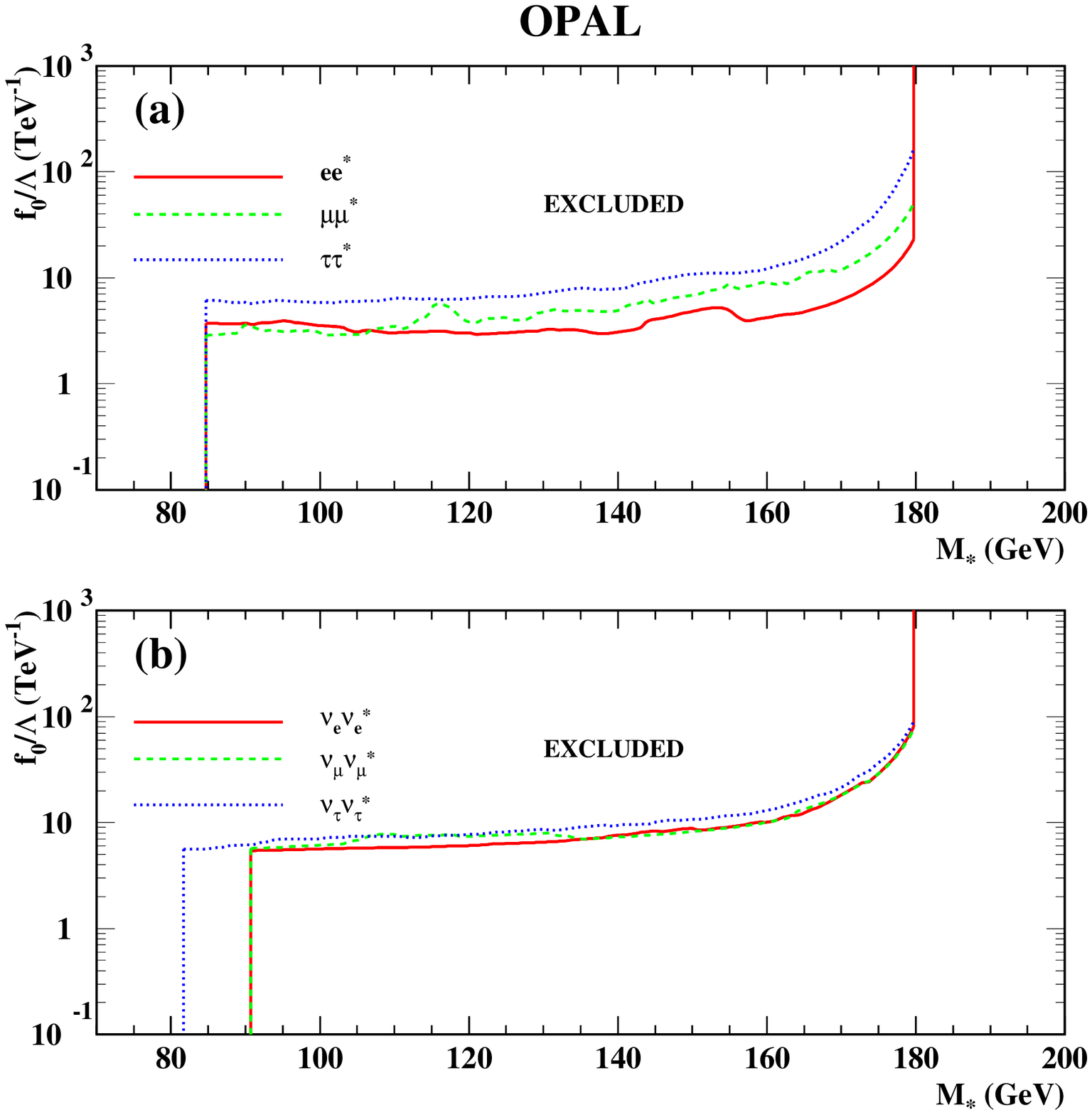}}
\caption{Upper limits at the 95\% CL on the ratio $f_0/\Lambda$
  of the coupling strength to the compositeness scale,
   as a function of the
  excited lepton mass.  (a) The limits on ${\rm e}^*$;
  $\mu^*$ and $\tau^*$ with arbitrary $\phi_f$;
  (b) the limits on  ${\rm \nu_e}^*$, ${\nu_\mu}^*$ and ${\nu_\tau}^*$ with arbitrary $\phi_f$.
  The regions above and to the left of the
  curves are excluded by the single- and pair-production searches,
  respectively.
  The symbol ${\rm M_*}$ represents the mass of the excited lepton.
  }
\label{f:elcouplim_f0}
\end{figure}

Finally, the results from different decay channels are combined:
\begin{equation}
L_{\lstar}\bigl(\frac{f_0^2}{\Lambda^2}\bigr) =
  L_{\lstar \rightarrow \ell \gamma }\bigl(\frac{f_0^2}{\Lambda^2}\bigr) \otimes
  L_{\lstar \rightarrow \ell \rm Z }\bigl(\frac{f_0^2}{\Lambda^2}\bigr) \otimes
  L_{\lstar \rightarrow \nu_\ell \rm W }\bigl(\frac{f_0^2}{\Lambda^2}\bigr) \ \ .
\end{equation}
The resulting likelihood functions from different decay topologies
are combined and upper limits on $\frac{f_0^2}{\Lambda^2}$ are inferred
as a function of the excited lepton mass by using the most conservative
limit as a function of $\phi_f$. 
The limit on $\frac{f_0^2}{\Lambda^2}$ for any value
of $\phi_f$ is determined and shown in Figure~\ref{f:elcouplim_f0}.
Using a similar technique, from the searches for
the pair-production of excited leptons,
the mass limits independent of the values of $f$ and $f^\prime$
are given in the last section of Tables~\ref{t:lstar_masslim} and 
~\ref{t:nustar_masslim} for charged and neutral excited leptons, respectively.
Together these represent the first limits on the compositeness scale
$f_0/\Lambda$
which do not depend on the relative values of $f$ and $\fp$.

%============================================================================
\section{Conclusion}
\label{s:conclusion}

We have searched for the
production of unstable heavy and excited leptons in
a data sample corresponding to an integrated luminosity of
$58~{\rm pb}^{-1}$ at a centre-of-mass energy of 181-184~GeV,
collected with the OPAL detector at LEP.
No evidence for their existence was found.
From the search for the pair-production of heavy and excited leptons,
lower mass limits were determined.
From the search for the single-production of excited leptons,
upper limits on $\sigma \times {\rm BR}$ of
${\rm e^+e^-} \rightarrow \ell^* \bar{\ell}, ~\ell^* \rightarrow \ell V$
and upper limits on the ratio
of the coupling to the compositeness scale were derived.
These limits supersede the results in~\cite{ref:hlOPAL172}.
Limits on the masses of excited leptons and on the compositeness
scale $f_0/\Lambda$ are established
independent of the relative values of
the coupling constants $f$ and $f^\prime$.

%============================================================================
\par
%\section*{Acknowledgements}
{\Large {\bf Acknowledgements}}
\par
\label{s:acknow}
We particularly wish to thank the SL Division for the efficient operation
of the LEP accelerator at all energies
 and for their continuing close cooperation with
our experimental group.  We thank our colleagues from CEA, DAPNIA/SPP,
CE-Saclay for their efforts over the years on the time-of-flight and trigger
systems which we continue to use.  In addition to the support staff at our own
institutions we are pleased to acknowledge the  \\
Department of Energy, USA, \\
National Science Foundation, USA, \\
Particle Physics and Astronomy Research Council, UK, \\
Natural Sciences and Engineering Research Council, Canada, \\
Israel Science Foundation, administered by the Israel
Academy of Science and Humanities, \\
Minerva Gesellschaft, \\
Benoziyo Center for High Energy Physics,\\
Japanese Ministry of Education, Science and Culture (the
Monbusho) and a grant under the Monbusho International
Science Research Program,\\
Japanese Society for the Promotion of Science (JSPS),\\
German Israeli Bi-national Science Foundation (GIF), \\
Bundesministerium f\"ur Bildung, Wissenschaft,
Forschung und Technologie, Germany, \\
National Research Council of Canada, \\
Research Corporation, USA,\\
Hungarian Foundation for Scientific Research, OTKA T-029328, 
T023793 and OTKA F-023259.\\

%============================================================================

%============================================================================

%%%%%%%%%%%%%%%%%%%%%%%%%%%%%%%%%%%%%%%%%%%%%%%%%%%%%%%%%%%%%%%%%

%%%%%%%%%%%%%%%%%%%%%%%%%%%%%%%%%%%%%%%%%%%%%%%%%%%%%%%%%%%%%%%%%

%%%%%%%%%%%%%%%%%%%%%%%%%%%%%%%%%%%%%%%%%%%%%%%%%%%%%%%%%%%%%%%%%

%%%%%%%%%%%%%%%%%%%%%%%%%%%%%%%%%%%%%%%%%%%%%%%%%%%%%%%%%%%%%%%%%

%%%%%%%%%%%%%%%%%%%%%%%%%%%%%%%%%%%%%%%%%%%%%%%%%%%%%%%%%%%%%%%%%

%%%%%%%%%%%%%%%%%%%%%%%%%%%%%%%%%%%%%%%%%%%%%%%%%%%%%%%%%%%%%%%%%

%%%%%%%%%%%%%%%%%%%%%%%%%%%%%%%%%%%%%%%%%%%%%%%%%%%%%%%%%%%%%%%%%

%============================================================================

\begin{thebibliography}{99}

\bibitem{ref:sm}
S.L. Glashow, J. Iliopoulos, and L. Maiani, Phys. Rev. {\bf D2} (1970) 1285;\\
S. Weinberg, Phys. Rev. Lett. {\bf 19} (1967) 1264;\\
A. Salam, {\em Elementary Particle Theory}, ed. N. Svartholm
(Almquist and Wiksells, Stockholm, 1968), 367.

\bibitem{ref:pdg} ``Review of Particle Physics''
R.M. Barnett {\it et al.}, Phys. Rev. D54 (1996).
  
\bibitem{ref:revue}
A.~Djouadi, D.~Schaile, C.~Verzegnassi, {\it et al.}, Report of the Working
Group ``Extended Gauge Models'' in
Proceedings of the Workshop ``${\rm e^+e^-}$ Collisions at 500 GeV: 
The Physics Potential'', P.~Zerwas, (ed.) Report DESY 92-123A+B.

\bibitem{ref:abdel} A. Djouadi, Z. Phys. C63 (1994) 317, and references therein.

\bibitem{ref:bdk} F.~Boudjema, A.~Djouadi, and J.L.~Kneur, Z.~Phys. C57 (1993) 425.


\bibitem{ref:hllep1}
ALEPH Collaboration, D.~Decamp   {\it et al.}, Phys. Lett. B236 (1990) 511; \\
OPAL Collaboration, M.Z.~Akrawy  {\it et al.}, Phys. Lett. B240 (1990) 250; \\
OPAL Collaboration, M.Z.~Akrawy  {\it et al.}, Phys. Lett. B247 (1990) 448; \\
L3 Collaboration, B.~Adeva       {\it et al.}, Phys. Lett. B251 (1990) 321; \\
DELPHI Collaboration, P.~Abreu   {\it et al.}, Phys. Lett. B274 (1992) 230.

\bibitem{ref:hlOPAL15}
OPAL Collaboration, G. Alexander {\it et al.}, Phys. Lett. B385 (1996) 433.

\bibitem{ref:hllep15}
L3 Collaboration,    M.~Acciarri {\it et al.}, Phys. Lett. B377 (1996) 304;  \\
ALEPH Collaboration, D.~Buskulic {\it et al.}, Phys. Lett. B384 (1996) 439.

\bibitem{ref:hlOPAL161}
  OPAL Collaboration, K.~Ackerstaff {\it et al.}, Phys. Lett. B393 (1997) 217.

\bibitem{ref:hlL3172}
  L3 Collaboration, M.~Acciarri {\it et al.}, 
  Phys. Lett. B412 (1997) 189.

\bibitem{ref:hlOPAL172}
 OPAL Collaboration, K.~Ackerstaff {\it et al.}, Eur. Phys. J. C1 (1998) 45.

\bibitem{ref:stableOPAL183}
 OPAL Collaboration, K.~Ackerstaff {\it et al.}, 
 Phys. Lett. B433 (1998) 195-208.

\bibitem{ref:DELPHI183}
 DELPHI Collaboration, P.~Abreu {\it et al.}, 
 Eur. Phys. J. C8 (1999) 41.

\bibitem{ref:ellep1}
OPAL Collaboration, M.Z.~Akrawy {\it et al.}, Phys. Lett. B257 (1990) 531;\\
ALEPH Collaboration, R~Barate {\it et al.},   Eur. Phys. J. C4 (1998) 571-590;\\
ALEPH Collaboration, D.~Decamp {\it et al.},   Phys. Lett. B250 (1990) 172;\\
DELPHI Collaboration, P.~Abreu {\it et al.},   Z.~Phys.    C53  (1992) 41; \\
L3 Collaboration, M.~Acciarri {\it et al.},    Phys. Lett. B353 (1995) 136.

\bibitem{ref:elopal15}
OPAL Collaboration, G.~Alexander {\it et al.}, Phys. Lett. B386 (1996) 463.

\bibitem{ref:ellep15}
L3 Collaboration, M.~Acciarri {\it et al.},   Phys. Lett. B370 (1996) 211;\\
DELPHI Collaboration, P.~Abreu {\it et al.},  Phys. Lett. B380 (1996) 480;\\
ALEPH Collaboration, D.~Buskulic {\it et al.},Phys. Lett. B385 (1996) 445.

\bibitem{ref:elopal161}
  OPAL Collaboration, K.~Ackerstaff {\it et al.}, Phys. Lett. B391 (1997) 197.
  
\bibitem{ref:ellep161}
  DELPHI Collaboration, P.~Abreu {\it et al.}, Phys. Lett. B393 (1997) 245;\\
  L3 Collaboration, M. Acciarri {\it et al.},  Phys. Lett. B401 (1997) 139.
  
\bibitem{ref:L3189}
  L3 Collaboration, M.~Acciarri {\it et al.}, 
  CERN-EP/99-138, October 1999, accepted by Phys.~Lett.~B.

\bibitem{ref:herasearches}
H1 Collaboration, I.~Abt {\it et al.}, Nucl. Phys. B396 (1993) 3;\\
ZEUS Collaboration, M.~Derrick {\it et al.}, Z. Phys. C65 (1994) 627; \\
H1 Collaboration, S. Aid {\it et al.}, Nucl. Phys. B483 (1997) 44.

\bibitem{ref:opalgg}
OPAL Collaboration, G.~Alexander {\it et al.}, Phys. Lett. B377 (1996) 222.

\bibitem{ref:2f} OPAL Collaboration, K.~Ackerstaff {\it et al.}, Phys. Lett. B391 (1997) 221;\\
  OPAL Collaboration, K.~Ackerstaff {\it et al.}, 
Eur. Phys. J. C2 (1998) 441

\bibitem{ref:OPAL-detector}
OPAL Collaboration, K.~Ahmet {\it et al.}, Nucl.~Instr.~Meth. A305 (1991) 275;\\
O.~Biebel {\it et al.},  Nucl.~Instr.~Meth. A323 (1992) 169 ;\\
M.~Hausschild {\it et al.},  Nucl.~Instr.~Meth. A314 (1992) 74;\\
%P.P.~Allport {\it et al.}, Nucl.~Instr.~Meth. A324 (1993) 34;\\
%P.P.~Allport {\it et al.}, Nucl.~Instr.~Meth. A346 (1994) 476;\\
B.E.~Anderson {\it et al.}, IEEE Trans.~Nucl.~Sci. 41 (1994) 845;\\
S.~Anderson  {\it et al.}, Nucl.~Instr.~Meth. A403 (1998) 326.

\bibitem{ref:superk} Super-Kamiokande Collaboration, Y.~Fukuda {\it et al.},
Phys.Rev.Lett. 81 (1998) 1562-1567.

\bibitem{ref:macro} MACRO Collaboration, M.~Ambrosio {\it et al.},
hep-ex/9908066, 31~Aug~1999.


\bibitem{ref:see-saw} M. Gell-Mann, P. Ramond, and R. Slansky,
Rev. Mod. Phys. 50 (1978) 721; \\
T. Yanagida, Phys. Rev. D20 (1979) 2986.

\bibitem{ref:nardi} E. Nardi, E. Roulet, and D. Tommasini, Phys. Lett. B344 (1995) 225.

\bibitem{ref:EXOTIC}
 R.~Tafirout and G.~Azuelos, "A  Heavy Fermion and Excited Fermion
       Monte Carlo Generator for ${\rm e^+e^-}$ Physics",
       accepted by Comp. Phys. Comm.

\bibitem{ref:zerwas} J.H.~K\"{u}hn, A.~Reiter, and P.M.~Zerwas, Nucl. Phys. B272 (1986) 560.

\bibitem{ref:jetset} T. Sj\"{o}strand, Comp.~Phys.~Comm. 82 (1994) 74.

\bibitem{ref:grc4f} 
  J.~Fujimoto {\it et al.}, Comp. Phys. Comm. 100 (1996) 128.

\bibitem{ref:phojet} R.~Engel and J.~Ranft, Phys. Rev. D54 (1996) 4244.

\bibitem{ref:herwig} G.~Marchesini {\it et al.}, Comp.~Phys.~Comm. 67 (1992) 465.

\bibitem{ref:bhwide}
S.~Jadach, W.~Placzek, and B.F.L.~Ward, Phys. Lett. B390 (1997) 298.

\bibitem{ref:teegg} D.~Karlen, Nucl. Phys. B289 (1987) 23.

\bibitem{ref:koralz}
S.~Jadach, B.F.L.~Ward, and Z.~W\c{a}s, Comp.~Phys.~Comm. 79 (1994) 503.

\bibitem{ref:vermaseren}
R.~Bhattacharya, J.~Smith, and G.~Grammer, Phys. Rev. D15 (1977) 3267;\\
J.~Smith, J.A.M.~Vermaseren, and G.~Grammer, Phys. Rev. D15 (1977) 3280.

\bibitem{ref:radcor}
F.A.~Berends and R.~Kleiss, Nucl. Phys. B186 (1981) 22.

%\bibitem{ref:nunugpv}
%  G. Montagna {\it et al.}, Nucl. Phys. B452 (1995) 161.

\bibitem{ref:gopal}
J.~Allison {\it et al.}, Nucl.~Instr.~Meth. A317 (1992) 47.


\bibitem{ref:durham} 
  N.~Brown and W.J.~Stirling, Phys. Lett. B252 (1990) 657; \\
  S.~Bethke, Z.~Kunszt, D.~Soper, and W.J.~Stirling, Nucl. Phys. B370 (1992) 310; \\
  S.~Catani {\it et al.}, Nucl. Phys. B269 (1991) 432; \\
  N.~Brown and W.~J.~Stirling, Z. Phys. C53 (1992) 629.

\bibitem{ref:gce}
  OPAL Collaboration, M.Z.~Akrawy {\it el al.}, Phys. Lett. B253 (1990) 511.

\bibitem{ref:pr258}
  OPAL Collaboration, G. Abbiendi et al, Eur. Phys. J. C8 (1999) 23.

\bibitem{ref:lyons} L. Lyons, {\it Statistics for Nuclear and Particle Physicists}, Cambridge University Press, Cambridge (1986).

\bibitem{ref:LL} OPAL Collaboration, G.~Alexander {\it et al.}, Z.~Phys. C52 (1991) 175.

\bibitem{ref:dedx} M. Hauschild {\it et al.}, Nucl. Instr. and Meth. A314 (1992) 74.

\bibitem{ref:NN}
  OPAL Collaboration, R.~Akers {\it et al.}, Phys. Lett. B327 (1994) 411.

\bibitem{ref:MU}
  OPAL Collaboration, P. Acton {\it et al.},
  Z.~Phys. C58 (1993) 523.

\end{thebibliography}
\end{document}